\documentclass[10pt,letterpaper]{article}
\usepackage[dvipsnames]{xcolor}
\usepackage[top=0.75in,left=1.3in,footskip=0.75in]{geometry}
\usepackage{amsfonts}
\usepackage{amsmath}
\usepackage[round]{natbib}
\usepackage{enumitem}
\usepackage[section]{placeins}
\usepackage[all]{xy}
\usepackage{url,tikz}
\usetikzlibrary{trees}
\usetikzlibrary{arrows,shapes,decorations,automata,backgrounds,petri}
\usepackage{bm}
\usepackage{hyperref}
\usepackage[right]{lineno}
\usepackage[title]{appendix}
\usepackage{titlesec}
\usepackage{booktabs}

\raggedright
\usepackage{lastpage,fancyhdr,graphicx}
\usepackage{epstopdf}
\newcommand{\ra}[1]{\renewcommand{\arraystretch}{#1}}
\fancyheadoffset[L]{1in}
\fancyfootoffset[L]{1in}
\newcommand{\wR}{\widehat{\mathcal{R}}}
\newcommand{\wW}{\mathcal{W}}
\newcommand{\be}{\bm{\varepsilon}}

\usepackage{comment}

\begin{document}

\begin{flushleft}
{\Large
\textbf{Outbreak dynamics and population vulnerability in stochastic epidemic models on networks}
}
\newline
\\
Makoto Ueki\textsuperscript{1},
Robin N. Thompson\textsuperscript{2},
Murad Banaji\textsuperscript{3*}

\bigskip
\textbf{1} School of Physics, Georgia Institute of Technology, Atlanta, GA, USA.
\\
\textbf{2} Mathematical Institute, University of Oxford, UK.
\\
\textbf{3} School of Mathematical Sciences, Lancaster University, UK.
\\
\bigskip

* m.banaji@lancaster.ac.uk

\end{flushleft}

\section*{Abstract}
  During infectious disease epidemics, pathogen transmission occurs in host populations made up of interacting subpopulations. Using stochastic simulation and analytical approximations, we examine how outbreak sizes in networked populations depend on network architecture, subpopulation sizes and the strength of coupling between subpopulations. We find, as expected, that mean outbreak sizes are frequently lower in networked populations than in homogeneous populations with the same basic reproduction number. However, after an outbreak ends, a networked population is often vulnerable to further outbreaks, and the ending of an outbreak may not imply herd immunity in any sense. Another key finding is that a relatively small amount of randomly distributed prior immunity can be more protective in a networked population than a homogeneous population, a phenomenon which can be reproduced analytically in certain cases. We also find that in networked populations, randomly distributed prior immunity is often more protective than infection-acquired immunity; but this conclusion can be reversed in populations with highly variable susceptibility. All of these conclusions have implications for designing outbreak control strategies that aim to reduce pathogen transmission during epidemics.
  
\vspace{0.5cm}
\noindent

{\bf Keywords.} herd immunity, compartmental epidemiological models, metapopulation models, spatial models, networks, outbreak size

\section*{Author summary}

During infectious disease epidemics, outbreak sizes depend not just on the transmissibility of the pathogen, but also on the structure of the host population and the distribution of prior immunity within it. In this paper, we examine how outbreak sizes in stochastic infectious disease models depend on the population network structure and the extent and distribution of prior immunity. We observe that outbreak sizes are generally lower in networked populations than in homogeneous populations with the same basic reproduction number, but that in networked populations outbreaks often end without herd immunity, leaving the population vulnerable to further outbreaks. We also find that a relatively small amount of randomly distributed prior immunity can reduce outbreak sizes in a networked population more than in a homogeneous population. Moreover, in contrast to some previous work, we find that random immunity can be more protective than immunity from prior infections in networked populations, although this conclusion is dependent on network architecture, subpopulation sizes and the strength of coupling. We conclude that outbreak control strategies are likely to be most effective when based on modelling that takes into account the network structure of the population in which the pathogen is spreading.


\section{Introduction}

During the COVID-19 pandemic, epidemiological modelling was used to demonstrate that in heterogeneous populations (i) relatively small disease outbreaks can lead to herd immunity; and/or (ii) infection-acquired immunity can provide greater population protection than vaccination (with vaccination modelled as randomly allocated immunity); see \cite{BrittonCOVID,Montalban,GomesJTB} for example. The intuition behind the latter claim is that individuals or subpopulations most likely to contract and spread disease tend to be infected early in outbreaks, rapidly reducing the vulnerability of the population as a whole to further outbreaks, a phenomenon termed ``network frailty'' by \cite{Ferrari}.

Our goal here is to explore the limits of these observations by analysing stochastic metapopulation models composed of subpopulations networked in various architectures. The framework includes individual-based models as an extreme case where each subpopulation consists of a single individual. Conversely, all models in our framework could be recast as {\em weighted} individual-based models (see \cite{Eames2009,Goltsev2012} for example, and \cite[Section VII.B.3]{PastorSantorras2015} for a general discussion of such networks). On the other hand, unweighted individual-based models constitute a strictly smaller class of models than those we consider here, and it may not be possible to reproduce some of the behaviours we describe here in unweighted individual-based models.

It is conceptually helpful to think of the subpopulations as spatial units: transmission between subpopulations could then reflect interactions between geographically separated subpopulations \citep{Riley:2007}, and would be strongly affected by interventions which reduce mobility, as applied, for example, during the COVID-19 pandemic. We remark that our choice to focus on metapopulation models stems in part from their potential for parameterisation from data (see, for example, \cite{Ferguson2001,Beyer2011,Pei2018,Iyaniwura2023}). Henceforth, we refer to all the models we study simply as {\em spatial models}, and the underlying populations as {\em networked populations}. 

The huge range of possible network structures, and the complex relationship between network properties and outbreak dynamics (see \cite{PastorSantorras2015} and the references therein), mean that extracting any universal principles relating the two is complicated. Our goal here is to extract some broad qualitative features of how population vulnerability to infection depends on network architecture, levels of transmission between subpopulations, subpopulation sizes and levels of prior immunity, whether arising via random immunisation, or via the spread of infection. 

In order to obtain meaningful conclusions, we require a metric which serves to quantify population vulnerability. We begin by observing that the notions of the {\em basic reproduction number} and {\em herd immunity} \citep{FineHerdImmunity} are frequently used to assess the extent to which a pathogen is likely to spread in a host population; however, a number of definitions of these quantities are possible, and their usefulness can depend on the context. The most natural definitions, in terms of the next-generation matrix \citep{DriesscheR0}, or equivalently in terms of the mean matrix \citep[Chapter 5]{AthreyaNey} as discussed in Appendix~\ref{appbranching}, are useful when subpopulations are large, but can have limitations. Below we discuss these limitations, and choose the {\em mean outbreak size following a random introduction} as a useful continuous measure which allows us to compare population vulnerability to disease across a range of model types and contexts. 

With this choice in place, we investigate how the mean outbreak size depends on the network architecture, subpopulation sizes, coupling strength and prior immunity in networked populations. We focus on demonstrating various qualitative features of these relationships, rather than attempting an exhaustive quantitative study. When connectivity of the network is low or coupling weak, we find that outbreak size distributions need not be strongly bimodal, and outbreaks may subside following relatively few infections, leaving populations vulnerable to further outbreaks. Thus, we should not, in general, infer herd immunity simply because outbreaks end. 

Another observation of real-world relevance in terms of the potential impacts of interventions is that even when a spatial model has similar expected outbreak size to its homogeneous counterpart, random immunisation (naively modelling vaccination) can have a stronger protective effect in the spatial case. Effectively, random immunisation in networked populations can magnify stochastic extinction, leading to a more rapid reduction in expected outbreak sizes. This contrasts with the observation in \cite[Chapter 12]{AndersonMay1991} that heterogeneity tends to {\em decrease} the effectiveness of random immunisation, with an extreme case being the absence of any herd immunity threshold under random immunisation in certain classes of networks \citep{PastorSantorras2002}.

A third observation is that, in many classes of spatial models, random immunisation can be {\em more protective} than infection-acquired immunity, leading to smaller expected secondary outbreaks. Thus, as also recently observed by \cite{Hiraoka2025}, the finding in some previous publications such as \cite{BrittonCOVID} that infection-acquired immunity more effectively reduces population vulnerability than random immunity depends sensitively on assumptions about network structure and can be reversed.

\subsection{Outline of the methodology}

Within each subpopulation in a networked population, we have a standard epidemic model which, for simplicity, we take to be an SIR model. We hold the values of basic epidemiological parameters constant between subpopulations. Additionally, we allow coupling, namely, transmission, between adjacent subpopulations. Our approach is related to standard approaches to reaction-diffusion processes on networks (\cite{Colizza2007a,Juher09,Ripoll15} for example), but differs in that we do not explicitly introduce diffusion of individuals between subpopulations; instead we let {\em disease} diffuse between subpopulations while maintaining subpopulation sizes constant, see \cite[Chapter 12]{AndersonMay1991} for a similar approach. This simplifies both stochastic simulation, and some of the accompanying analysis.

Our main tool for studying epidemics in spatial models is extensive stochastic simulation of the associated continuous-time Markov chains (CTMCs) using Gillespie's direct stochastic simulation algorithm \citep{Gillespie}. This allows us to observe and analyse model behaviours which are fundamentally stochastic in nature, and would not be observed in the corresponding deterministic models. All code underlying the simulation results is available on GitHub \citep{muradEpigithub}. In some simple cases, especially when subpopulations are large and coupling is weak, we are able to compare the results of our simulations against formulae derived using branching-process approximations \citep{AllenMBI,AthreyaNey}. These calculations also allow us to provide some intuition to help explain the behaviours we observe in simulations.

\subsection{Summary of conclusions, and the underlying intuition}

Our main conclusions about network-based epidemic models are the following.
\begin{enumerate}[align=left,leftmargin=*,itemsep=4pt]
\item {\em Outbreak sizes depend on both network architecture and coupling strength.} As might be expected, more global interactions and/or stronger coupling lead to larger mean outbreak sizes. These findings are robust, although the qualitative features of the relationship between network architecture/coupling and mean outbreak size vary. We will see that, in addition to the mean, other features of the outbreak size distribution are strongly dependent on network architecture and coupling strength. Especially important to questions of herd immunity is the extent to which this distribution is bimodal, allowing us to distinguish ``minor'' from ``major'' outbreaks.

\item {\em Outbreaks often end without herd immunity.} In particular, when a pathogen becomes extinct, the population may be left in a state in which it is vulnerable to further outbreaks. 
  In the extreme case of a population composed of {\em disconnected} subpopulations, this claim becomes obvious: there is no contagion between subpopulations, and outbreaks are confined to the subpopulations in which infection is introduced, resulting in low outbreak sizes, but without much subsequent population protection against a random introduction. Regardless of network architecture. weak coupling, leading to low probabilities of contagion between subpopulations, does not qualitatively change these conclusions: most outbreaks remain {\em local}. Even in strongly-coupled models with relatively small subpopulations and low connectivity, outbreaks can remain largely local: spatial structure magnifies stochastic extinction effects.
  
\item {\em Random immunity can be more protective than in the corresponding homogeneous population.} Relatively low levels of random immunisation in a spatially structured population can lower outbreak sizes more than the same level of random immunisation in a homogeneous population. However, the opposite can also occur in some circumstances. The intuition behind this claim is that decreasing the susceptible fraction of the population evenly across subpopulations not only decreases the expected outbreak size in each subpopulation where infection arrives; it also decreases the {\em probability of contagion} between subpopulations. The latter effect can cause mean outbreak sizes to decrease more rapidly as we add random immunity to a spatially structured population than to a homogeneous population. We explore this finding with analysis and simulation, and find that while it holds across many models it can also be reversed for some architectures and sufficiently strong coupling.

\item {\em Random immunity can be more protective than infection-acquired immunity.} This holds frequently, but can be reversed for certain choices of network architecture, coupling, and subpopulation sizes. Whether random or infection-acquired immunity is more protective depends on competing effects. While network frailty means that infection-acquired immunity can break the most substantial pathways of transmission, random immunisation can uniformly reduce the probability of contagion between subpopulations by magnifying stochastic extinction. Which of these effects dominates depends on details of the network. Taking mean outbreak size as a measure of vulnerability, we find that prior random immunity can often be more protective than infection-acquired immunity; but this conclusion can be reversed when subpopulations are sufficiently small and there is sufficient variability in the susceptibility of individuals in different subpopulations. Interestingly, variability in infectivity is {\em not} sufficient to ensure that network frailty effects dominate, namely, susceptibility and infectivity play rather different roles when it comes to the spread of epidemics on networks.

\end{enumerate}

All the claims highlight that, in any particular real-world context, arriving at reliable conclusions about the protective effects of immunity, acquired through vaccination or infection and distributed in various ways within the population, would require detailed modelling of the population structure and pathogen dynamics relevant to that context.

\section{Methods}
\subsection{Model set-up and basic definitions}

The population is divided into $n$ subpopulations, which we assume at the outset have identical size $N_c$, and population density normalised to be $1$ (i.e., units of volume are chosen so each spatial unit has volume $N_c$). We define $N:= nN_c$ to be the total population. We allow $N_c=1$, in order to allow {\em individual-based models}. The dynamics follows the reaction scheme
\begin{equation}
  \label{mainreacscheme}
  \mathsf{I}_i + \mathsf{S}_j \overset{\varepsilon_{ij}\beta}{\longrightarrow} \mathsf{I}_i + \mathsf{I}_j,\quad \mathsf{I}_i \overset{\gamma}{\longrightarrow} \mathsf{R}_i \quad (i,j=1,\ldots,n)\,.
\end{equation}
Here, $\varepsilon_{ij}\beta$ and $\gamma$ are the deterministic mass action rate constants for the corresponding reactions, and $\mathsf{S}_i$, $\mathsf{I}_i$ and $\mathsf{R}_i$ denote, respectively, the susceptible, infected and recovered individuals in subpopulation $i$. Note that we allow transmission between subpopulations, but do not explicitly introduce migration of individuals, thus maintaining subpopulation sizes constant. 

The topology of the network is captured in the $n \times n$ {\em coupling matrix} $\be = (\varepsilon_{ij})$, which is normalised by setting $\sum_{i,j=1}^n\varepsilon_{ij} = n$ (or in Gamma networks, see below, setting the expected value of $\sum_{i,j=1}^n\varepsilon_{ij}$ to be $n$). We refer to the parameter $\varepsilon_{ij}$ ($i \neq j$) as the {\em leak} from subpopulation $i$ to subpopulation $j$. 
The uncoupled system corresponds to $\be$ being the identity matrix, while weak coupling corresponds to $\be$ being close to the identity matrix. Thus, in the absence of coupling, disease dynamics are identical in all subpopulations. Equation~\ref{mainreacscheme} defines ordinary differential equation models and CTMC models in the standard way, see for example \cite{Erban_Chapman_2020,AndersonKurtz,Allenprimer,AllenIntro}.

Let $S_i, I_i$ and $R_i$ denote the number of individuals in each state in subpopulation $i$, implicitly assumed to depend on time $t$, so that $S_i+I_i+R_i = N_c$. The stochastic propensity of the infection reaction $\mathsf{I}_i + \mathsf{S}_j \longrightarrow \mathsf{I}_i + \mathsf{I}_j$ is $\varepsilon_{ij} \beta I_iS_j/N_c$, while for the recovery reaction $\mathsf{I}_i \longrightarrow \mathsf{R}_i$ it is $\gamma I_i$.

\subsection{Next generation matrix, contagion probabilities, infectivity and susceptibility}
\label{secnextgen}
We let $\wW$ denote the transpose of the next-generation matrix \citep{DriesscheR0} corresponding to the reaction scheme (\ref{mainreacscheme}) above. We calculate, in a standard way, that
\begin{equation}
  \label{eqW}
\wW_{ij} = \varepsilon_{ij}\frac{\beta}{\gamma} \frac{S_j}{N_c}\, \quad \mbox{so that,} \quad \wW = \frac{\beta}{\gamma N_c}\, \be\, \mathrm{diag}(S)
\end{equation}
where $\mathrm{diag}(S)$ is a diagonal matrix with diagonal entries $S_i$. Note that the quantities $\wW_{ij}$, hence $\wW$, depend on $t$ via $S_j$ and $S$ respectively, but we have suppressed this dependence for readability. In the stochastic setting, $\wW$ tells us about {\em contagion probabilities}: under the branching process approximation (see, for example, \cite{AllenMBI} or Chapter 5 of \cite{AthreyaNey}), we can compute the probability that an infected individual in subpopulation $i$ recovers without (directly) infecting a susceptible individual in subpopulation $j$ to be
\begin{equation}
  \label{eqcontagion}
p_{ij}:=\frac{\gamma}{\gamma + \varepsilon_{ij}\beta S_j/N_c}  = \frac{1}{1 + \frac{\varepsilon_{ij}\beta S_j}{\gamma N_c}}  = \frac{1}{1+\wW_{ij}}\,.
\end{equation}
Details are given in Appendix~\ref{appbranching}. More generally, the number of secondary infections in subpopulation $j$ triggered by a single infection in subpopulation $i$ is geometrically distributed with parameter $p_{ij}$, and so the mean number of secondary infections in subpopulation $j$ directly triggered by a single infection in subpopulation $i$ is precisely $\wW_{ij}$.

Define $\mathcal{R}^{(i)}_t$ to be the expected number of secondary infections generated by a new infection in subpopulation $i$ at time $t$. Under the branching process approximation, we obtain:
\begin{equation}
\mathcal{R}^{(i)}_t =  \sum_{j=1}^n\wW_{ij} = \frac{\beta}{\gamma N_c}\sum_{j=1}^n\varepsilon_{ij}S_j\,, \quad \mbox{so that } \quad \mathcal{R}_0^{(i)} = \frac{\beta}{\gamma}\sum_{j=1}^n\varepsilon_{ij}\,,
\end{equation}
can be seen as the {\em infectivity} of individuals in the $i$th subpopulation. As $\mathcal{R}_0^{(i)}$ is proportional to the $i$th row-sum of $\be$, we refer to models where row-sums of $\be$ are constant as {\em constant-infectivity} models. Throughout this paper, ``constant-infectivity'' will always mean that row-sums of $\be$ are all equal to $1$, i.e., $\be$ is row-stochastic and $\mathcal{R}_0^{(i)} = \beta/\gamma$ for all $i$.

Define $C^{(j)}_t$ to be the expected number of infections (directly) triggered in subpopulation $j$ by a {\em random introduction}, namely, an infection introduced into each subpopulation with equal likelihood. Under the branching process approximation, this is 
\begin{equation}
C^{(j)}_t:= \frac{1}{n}\sum_{i=1}^n \wW_{ij} = \frac{\beta}{\gamma} \frac{S_j}{N}\sum_{i=1}^n \varepsilon_{ij}\,, \quad \mbox{so that} \quad  C_0^{(j)} = \frac{\beta}{\gamma n}\sum_{i=1}^n \varepsilon_{ij}
\end{equation}
can be seen as the {\em susceptibility} of individuals in the $j$th subpopulation. As $C_0^{(j)}$ is proportional to the $j$th column-sum of $\be$, we refer to models where $\be$ has constant column-sums as {\em constant-susceptibility} models. Throughout this paper, ``constant-susceptibility'' will always mean that column-sums of $\be$ are all equal to $1$, and $C_0^{(j)} = \beta/(\gamma n)$ for all $j$.

Many of the models we examine will have neither constant infectivity nor constant susceptibility; but in order to better understand the behaviours we observe, we will sometimes rescale rows or columns of $\be$ to obtain constant-infectivity or constant-susceptibility models. 

\subsection{Reproduction numbers and herd immunity} The effective reproduction number is often defined as {\em the expected number of secondary infections resulting directly from a primary infection} \citep{GosticReff, metcalf2015understanding, nishiura2009effective}. For this to be well-defined in a spatial model, we need to assign a probability that a primary infection occurs in a given subpopulation. Here, we make the assumption of a primary infection in a randomly chosen subpopulation. With this assumption, in the large population limit, we obtain the {\em classical effective reproduction number}
\begin{equation}
  \label{classicRt}
\wR_t:= \frac{1}{n}\sum_{i=1}^n \mathcal{R}^{(i)}_t = \frac{\beta}{\gamma N}\sum_{i,j=1}^n\varepsilon_{ij}S_j\, , \quad \mbox{so that} \quad \wR_0 = \frac{\beta}{\gamma}\,\frac{1}{n}\sum_{i,j=1}^n\varepsilon_{ij}\,.
\end{equation}
$\wR_0$ is a key parameter: throughout this work, when comparing outbreak sizes in any set of spatial models, we hold expected or achieved $\wR_0$ constant across the different models. For constant-susceptibility networks the classical effective reproduction number simplifies to:
\begin{equation}
  \label{RtclassA}
\wR_t = \frac{\beta}{\gamma N}\sum_{i,j=1}^n\varepsilon_{ij}S_j = \frac{\beta S_{tot}}{\gamma N} = \wR_0 \frac{S_{tot}}{N}\,,
\end{equation}
where $S_{tot} := \sum_{j=1}^nS_j$ is the total susceptible population. 

We remark that the theoretical definition of $\wR_t$ used here will, in general, differ from empirical values of $\wR_t$ estimated from real-world data during an outbreak. The latter would provide estimates of the number of secondary infections resulting from each actually occurring primary infection, rather than from a hypothetical random infection.

Define $\rho(A)$ to be the spectral radius of a square matrix $A$. Following \cite{DriesscheR0}, we define the {\em next-generation effective reproduction number} to be
\begin{equation}
  \label{nextgenRt}
  \mathcal{R}_t:= \rho(\wW^\top) = \rho(\wW)\,, \quad \mbox{so that} \quad \mathcal{R}_0 = \frac{\beta}{\gamma}\rho(\be)\,.
\end{equation}
Thus for networks of constant infectivity or constant susceptibility,
\begin{equation}
  \mathcal{R}_0 = \frac{\beta}{\gamma} = \wR_0\,.
\end{equation}

In inhomogeneous populations, defining {\em herd immunity} presents both practical difficulties (see, for example, \cite{Morens} in the context of the COVID-19 pandemic), and theoretical difficulties. In spatial models, provided outbreak size distributions are strongly bimodal, the {\em herd immunity threshold} can be defined heuristically as the boundary between two regions of the state space, in one of which (projecting onto the disease-free states) the probability of a ``major outbreak'' is zero, while in the other it is positive. This is easily formalised mathematically as the set of states at which, $\mathcal{R}_t = 1$ or, equivalently, in terms of the spectral radius of the mean matrix from multi-type branching process theory (details of the equivalence are in Appendix~\ref{appbranching}). Note, however, that these derivations implicitly assume large subpopulation sizes and that, in spatial models, outbreak probabilities and/or expected outbreak sizes can become small even when we are far from the theoretical herd immunity threshold (see Section~\ref{secprotect} below). In Section~\ref{secprotect} we discuss a practical approach in terms of expected outbreak sizes to defining herd immunity in network models with finite subpopulations.

\subsection{Contagion probabilities for weak coupling}
\label{seccontagion1}
If $\wW_{ij} \ll 1$ for some $i \neq j$, we can approximate the geometric distribution of secondary infections in subpopulation $j$ directly caused by a primary infection in subpopulation $i$ with a Bernoulli distribution, i.e.,
\begin{equation}
\mathbb{P}(\mbox{$0$ secondary infections in $j$}) \approx 1-\wW_{ij}, \quad \mathbb{P}(\mbox{$1$ secondary infection in $j$}) \approx \wW_{ij}\,,
\end{equation}
with $\mathbb{P}(\mbox{$k$ secondary infections in $j$})$ negligible for $k > 1$. Let us assume that $\wW_{jj}>1$, i.e., there is a probability of approximately $1-1/\wW_{jj}$ that an introduction into subpopulation $j$ directly triggers an outbreak in subpopulation $j$ \citep{Southall}. Then, given $m$ primary infections in subpopulation $i$, the probability of $k$ secondary infections in subpopulation $j$ {\em and no outbreak} in subpopulation $j$ is approximately 
\begin{equation}
p_k^{(ij)}(m) := \left(\frac{1}{\wW_{jj}}\right)^k{m \choose k}(1-\wW_{ij})^{m-k}(\wW_{ij})^{k} = {m \choose k}(1-\wW_{ij})^{m-k}\left(\frac{\wW_{ij}}{\wW_{jj}}\right)^{k}\,.
\end{equation}
The probability of an outbreak of size $m$ in subpopulation $i$ {\em not} triggering an outbreak in subpopulation $j$ is then, approximately,
\begin{equation}
  \label{eqalpha1}
\alpha_{ij}(m) := \sum_{k=1}^mp_k^{(ij)}(m) = \left(1-\wW_{ij}+\frac{\wW_{ij}}{\wW_{jj}}\right)^m = \left(1-\wW_{ij}+\frac{\varepsilon_{ij}}{\varepsilon_{jj}}\right)^m = \alpha_{ij}(1)^m\,.
\end{equation}
The probability of direct contagion from subpopulation $i$ to subpopulation $j$, given $m$ infections in subpopulation $i$, is thus approximately $1-\alpha_{ij}(m) = 1-\alpha_{ij}(1)^m$. Note that provided $\wW_{ij}>0$ and $\wW_{jj}>1$, we have $\alpha_{ij}(1) < 1$, hence $\lim_{m \to \infty} \alpha_{ij}(m) = 0$, i.e., contagion occurs with probability $1$ in the limit of large $m$. 

We will use Equation (\ref{eqalpha1}) in Section~\ref{seccompvhom} below to explain why random immunity can be more protective in spatial models than in homogeneous models, even when typical outbreak sizes in the absence of prior immunity are similar. We will also use it to derive quantitative estimates of outbreak sizes in some special cases, and compare these with the results of simulation.

\subsection{The mean leak}
We introduce a parameter useful for controlling the coupling strength, namely the extent of transmission between subpopulations, in models with a variety of architectures. Define
\begin{equation}
\Theta_i := \sum_{j=1, j \neq i}^n\varepsilon_{ij} \quad \mbox{and} \quad \Theta := \frac{1}{n}\sum_{i=1}^n \Theta_i = \frac{1}{n}\sum_{i,j,i \neq j}\varepsilon_{ij} \,.
\end{equation}
$\Theta_i$ measures the level of transmission from subpopulation $i$ to its neighbours. We refer to $\Theta$ as the {\em mean leak} of the network. $\Theta$ can be interpreted as the fraction of secondary infections which are expected to occur between subpopulations (as opposed to within subpopulations) given a random introduction into a large, completely susceptible, population.

In individual-based networks, i.e., when $N_c=1$, we are forced to set $\Theta = 1$ as, clearly, no secondary infections can occur within a subpopulation. The reader may confirm that setting $N_c=1$ and $\Theta = 1$, we obtain precisely the CTMC we would expect for the corresponding individual-based model. In all other cases, $\Theta$ is a parameter we control, and when $\Theta \ll 1$ we say that the network is {\em weakly coupled}. 

\subsection{Network and coupling architectures}

We construct all networks, with the exception of Gamma networks (discussed separately below), as follows: (i) we choose an underlying (undirected) graph, and (ii) we use this graph to set entries in the coupling matrix $\be$. We draw on \cite{KeelingEames, Ferrari} for interesting network architectures to examine in an epidemiological context.

In order to set $\be$ in a manner consistent with the chosen network architecture, in all cases (including Gamma networks) we start by fixing $\Theta \in [0,1]$, and setting $\varepsilon_{ii} = 1-\Theta$ so that within-subpopulation dynamics is identical in all subpopulations. After this, for networks other than Gamma networks, we set $\varepsilon_{ij}=\frac{\Theta}{d}$ ($i \neq j$), where $d$ is the mean degree of the vertices. Thus, by default, all non-zero off-diagonal entries in $\be$ are identical. Leaving aside Gamma networks and individual-based networks, we get $\varepsilon_{ii} = \varepsilon_{ij}$ when $\Theta = \Theta_{max} := d/(d+1)$. We can thus take $\Theta=\Theta_{max}$ to be the {\em fully coupled} case for most of the networks we discuss here.

To define various classes of networks of interest, we need a couple of preliminary definitions. A {\em regular ring lattice} (or {\em ring} for short) with parameters $(n,K)$ is constructed from $n$ vertices arranged in a ring, by adding edges between each vertex and its $K<n$ nearest neighbours, where $K$ is assumed even. A {\em periodic square lattice} (or {\em grid} for short) is an $N_x \times N_y$ grid with each vertex adjacent to its four nearest neighbours, and with periodic boundary conditions (i.e., a torus).

The networks we consider are:
\begin{enumerate}[align=left,leftmargin=*,itemsep=4pt]
\item {\em Complete, symmetric networks.} The underlying graph is a complete graph. 
\item {\em Random networks.} We choose $\hat{d} < n-1$, the {\em expected mean degree}. Given each pair of distinct vertices $i \neq j$, the edge $(i,j)$ is included with probability $\hat{d}/(n-1)$ \citep{GilbertRandom}. 
\item {\em Small-world networks.} We choose $p \in [0,1]$, the {\em rewiring probability}.
  \begin{enumerate}
    \item[(i)] {\em Small-world rings.} starting with an $(n,K)$ regular ring lattice, we rewire according to the Watts-Strogatz procedure \citep{WattsStrogatz}: for each vertex $i$, for each vertex $j$ adjacent to $i$ and clockwise from $i$, with probability $p$ we choose a non-edge $(i,k)$ uniformly at random from current non-edges of this form, and replace edge $(i,j)$ with edge $(i,k)$. Note that the mean degree of the underlying graph remains $K$.
    \item[(ii)] {\em Small-world grids.} Starting from an $N_x \times N_y$ grid, we again rewire according to the Watts-Strogatz procedure. For each vertex $i$, for each vertex $j$ adjacent to $i$ to the right or below (bearing in mind the periodic boundary conditions), with probability $p$ we choose a non-edge $(i,k)$ uniformly at random from current non-edges of this form, and replace edge $(i,j)$ with edge $(i,k)$. Note that this preserves the mean degree of the graph at $4$.
  \end{enumerate}

  \item {\em Ring-preferential networks.} We choose parameters $1\leq n_r < n$ ({\em initial ring size}), $1 \leq K \leq n_r$ ({\em initial clique size}, which must be even), and $1 \leq d_0 \leq K$ ({\em attachment degree}, namely, the degree of new vertices added). We begin with an $(n_r,K)$ regular ring lattice and follow the {\em preferential attachment} procedure of Barab\'asi and Albert \citep{BarabasiAlbert}: we add vertices until we have a total of $n$ vertices, attaching each new vertex to $d_0$ distinct existing vertices chosen with probability proportional to their degree. This results in a network with mean degree $(n_rK + 2d_0(n-n_r))/n$, and (provided $n \gg n_r$) large variance in degrees.
        
  \item {\em Gamma networks.} For these networks $\Theta$ is the {\em expected} mean leak, and we choose a shape parameter $\alpha > 0$ which controls the variance of coupling strengths. There are two cases.
  \begin{enumerate}
  \item[(i)] {\em Asymmetric.} For each $(i,j)$, $i \neq j$, we choose $\varepsilon_{ij}$ independently from a Gamma distibution with shape $\alpha$ and rate $\alpha(n-1)/\Theta$ (hence, mean $\Theta/(n-1)$ and variance $\Theta^2/(\alpha(n-1)^2)$). 
  \item[(ii)] {\em Symmetric.} For each $(i,j)$, $i < j$, we choose each $\varepsilon_{ij}$ independently from a Gamma distibution with shape $\alpha$ and rate $\alpha(n-1)/\Theta$. We then set $\varepsilon_{ji} = \varepsilon_{ij}$. 
  \end{enumerate}
  A ``Gamma network'', without qualification, will mean an asymmetric Gamma network. Note that in Gamma networks, the network topology is captured directly in $\be$, rather than in the underlying graph: in this sense they are exceptional amongst the networks we consider in this paper. By construction, for Gamma networks, we know the expected, rather than exact, value of the mean leak, hence of $\wR_0$.
\end{enumerate}

The various network types are illustrated schematically in Figure~\ref{fignetworktypes}. 

\begin{figure}[ht]
  \begin{center}
    \underline{Schematic illustrations of the various network types}
\end{center}

\begin{center}
\begin{tikzpicture}[domain=0:14,scale=0.5]
\draw[very thin,color=white] (-4,-2.5) grid (4,2.5);

\node at (-3,2) {(a)};

\path (0: 2cm) coordinate (S1);
\path (40: 2cm) coordinate (S2);
\path (2*40: 2cm) coordinate (S3);
\path (3*40: 2cm) coordinate (S4);
\path (4*40: 2cm) coordinate (S5);
\path (5*40: 2cm) coordinate (S6);
\path (6*40: 2cm) coordinate (S7);
\path (7*40: 2cm) coordinate (S8);
\path (8*40: 2cm) coordinate (S9);

\fill (S1) circle (2pt);
\fill (S2) circle (2pt);
\fill (S3) circle (2pt);
\fill (S4) circle (2pt);
\fill (S5) circle (2pt);
\fill (S6) circle (2pt);
\fill (S7) circle (2pt);
\fill (S8) circle (2pt);
\fill (S9) circle (2pt);

\draw[thin,color=black] (S1) -- (S2);
\draw[thin,color=black] (S1) -- (S3);
\draw[thin,color=black] (S1) -- (S4);
\draw[thin,color=black] (S1) -- (S5);
\draw[thin,color=black] (S1) -- (S6);
\draw[thin,color=black] (S1) -- (S7);
\draw[thin,color=black] (S1) -- (S8);
\draw[thin,color=black] (S1) -- (S9);
\draw[thin,color=black] (S2) -- (S3);
\draw[thin,color=black] (S2) -- (S4);
\draw[thin,color=black] (S2) -- (S5);
\draw[thin,color=black] (S2) -- (S6);
\draw[thin,color=black] (S2) -- (S7);
\draw[thin,color=black] (S2) -- (S8);
\draw[thin,color=black] (S2) -- (S9);
\draw[thin,color=black] (S3) -- (S4);
\draw[thin,color=black] (S3) -- (S5);
\draw[thin,color=black] (S3) -- (S6);
\draw[thin,color=black] (S3) -- (S7);
\draw[thin,color=black] (S3) -- (S8);
\draw[thin,color=black] (S3) -- (S9);
\draw[thin,color=black] (S4) -- (S5);
\draw[thin,color=black] (S4) -- (S6);
\draw[thin,color=black] (S4) -- (S7);
\draw[thin,color=black] (S4) -- (S8);
\draw[thin,color=black] (S4) -- (S9);
\draw[thin,color=black] (S5) -- (S6);
\draw[thin,color=black] (S5) -- (S7);
\draw[thin,color=black] (S5) -- (S8);
\draw[thin,color=black] (S5) -- (S9);
\draw[thin,color=black] (S6) -- (S7);
\draw[thin,color=black] (S6) -- (S8);
\draw[thin,color=black] (S6) -- (S9);
\draw[thin,color=black] (S7) -- (S8);
\draw[thin,color=black] (S7) -- (S9);
\draw[thin,color=black] (S8) -- (S9);


\end{tikzpicture}
\begin{tikzpicture}[domain=0:14,scale=0.5]
\draw[very thin,color=white] (-4,-2.5) grid (4,2.5);
\node at (-3,2) {(b)};

\path (0: 2cm) coordinate (S1);
\path (30: 2cm) coordinate (S2);
\path (2*30: 2cm) coordinate (S3);
\path (3*30: 2cm) coordinate (S4);
\path (4*30: 2cm) coordinate (S5);
\path (5*30: 2cm) coordinate (S6);
\path (6*30: 2cm) coordinate (S7);
\path (7*30: 2cm) coordinate (S8);
\path (8*30: 2cm) coordinate (S9);
\path (9*30: 2cm) coordinate (S10);
\path (10*30: 2cm) coordinate (S11);
\path (11*30: 2cm) coordinate (S12);

\fill (S1) circle (2pt);
\fill (S2) circle (2pt);
\fill (S3) circle (2pt);
\fill (S4) circle (2pt);
\fill (S5) circle (2pt);
\fill (S6) circle (2pt);
\fill (S7) circle (2pt);
\fill (S8) circle (2pt);
\fill (S9) circle (2pt);
\fill (S10) circle (2pt);
\fill (S11) circle (2pt);
\fill (S12) circle (2pt);

\draw[thin,color=black] (S1) -- (S3);
\draw[thin,color=black] (S1) -- (S10);

\draw[thin,color=black] (S2) -- (S5);
\draw[thin,color=black] (S2) -- (S12);

\draw[thin,color=black] (S3) -- (S1);
\draw[thin,color=black] (S3) -- (S4);
\draw[thin,color=black] (S3) -- (S8);
\draw[thin,color=black] (S3) -- (S11);
\draw[thin,color=black] (S3) -- (S12);

\draw[thin,color=black] (S4) -- (S3);
\draw[thin,color=black] (S4) -- (S5);
\draw[thin,color=black] (S4) -- (S6);

\draw[thin,color=black] (S5) -- (S2);
\draw[thin,color=black] (S5) -- (S4);
\draw[thin,color=black] (S5) -- (S7);
\draw[thin,color=black] (S5) -- (S8);
\draw[thin,color=black] (S5) -- (S12);

\draw[thin,color=black] (S6) -- (S4);
\draw[thin,color=black] (S6) -- (S8);
\draw[thin,color=black] (S6) -- (S9);

\draw[thin,color=black] (S7) -- (S5);
\draw[thin,color=black] (S7) -- (S8);

\draw[thin,color=black] (S8) -- (S3);
\draw[thin,color=black] (S8) -- (S5);
\draw[thin,color=black] (S8) -- (S6);
\draw[thin,color=black] (S8) -- (S7);
\draw[thin,color=black] (S8) -- (S10);
\draw[thin,color=black] (S8) -- (S11);
\draw[thin,color=black] (S8) -- (S12);

\draw[thin,color=black] (S9) -- (S6);
\draw[thin,color=black] (S9) -- (S10);
\draw[thin,color=black] (S9) -- (S11);
\draw[thin,color=black] (S9) -- (S12);

\draw[thin,color=black] (S10) -- (S1);
\draw[thin,color=black] (S10) -- (S8);
\draw[thin,color=black] (S10) -- (S10);
\draw[thin,color=black] (S10) -- (S11);

\draw[thin,color=black] (S11) -- (S3);
\draw[thin,color=black] (S11) -- (S8);
\draw[thin,color=black] (S11) -- (S9);
\draw[thin,color=black] (S11) -- (S10);

\draw[thin,color=black] (S12) -- (S2);
\draw[thin,color=black] (S12) -- (S3);
\draw[thin,color=black] (S12) -- (S5);
\draw[thin,color=black] (S12) -- (S8);
\draw[thin,color=black] (S12) -- (S9);

\end{tikzpicture}
\begin{tikzpicture}[domain=0:14,scale=0.5]
\draw[very thin,color=white] (-4,-2.5) grid (4,2.5);
\node at (-3,2) {(c)};

\path (0: 2cm) coordinate (S1);
\path (30: 2cm) coordinate (S2);
\path (2*30: 2cm) coordinate (S3);
\path (3*30: 2cm) coordinate (S4);
\path (4*30: 2cm) coordinate (S5);
\path (5*30: 2cm) coordinate (S6);
\path (6*30: 2cm) coordinate (S7);
\path (7*30: 2cm) coordinate (S8);
\path (8*30: 2cm) coordinate (S9);
\path (9*30: 2cm) coordinate (S10);
\path (10*30: 2cm) coordinate (S11);
\path (11*30: 2cm) coordinate (S12);

\fill (S1) circle (2pt);
\fill (S2) circle (2pt);
\fill (S3) circle (2pt);
\fill (S4) circle (2pt);
\fill (S5) circle (2pt);
\fill (S6) circle (2pt);
\fill (S7) circle (2pt);
\fill (S8) circle (2pt);
\fill (S9) circle (2pt);
\fill (S10) circle (2pt);
\fill (S11) circle (2pt);
\fill (S12) circle (2pt);

\draw[thin,color=black] (S1) -- (S2);
\draw[thin,color=black] (S1) -- (S3);
\draw[thin,color=black] (S2) -- (S3);
\draw[thin,color=black] (S2) -- (S4);
\draw[thin,color=black] (S3) -- (S5);
\draw[thin,color=black] (S4) -- (S5);
\draw[thin,color=black] (S4) -- (S6);
\draw[thin,color=black] (S5) -- (S6);
\draw[thin,color=black] (S5) -- (S7);
\draw[thin,color=black] (S6) -- (S7);
\draw[thin,color=black] (S7) -- (S8);
\draw[thin,color=black] (S7) -- (S9);
\draw[thin,color=black] (S8) -- (S9);
\draw[thin,color=black] (S8) -- (S10);
\draw[thin,color=black] (S8) -- (S9);
\draw[thin,color=black] (S8) -- (S10);
\draw[thin,color=black] (S9) -- (S10);
\draw[thin,color=black] (S10) -- (S11);
\draw[thin,color=black] (S10) -- (S12);
\draw[thin,color=black] (S11) -- (S12);
\draw[thin,color=black] (S11) -- (S1);
\draw[thin,color=black] (S12) -- (S1);
\draw[thin,color=black] (S12) -- (S2);

\draw[thin,color=black] (S1) -- (S4);
\draw[thin,color=black] (S5) -- (S9);
\draw[thin,color=black] (S8) -- (S11);

\end{tikzpicture}
\end{center}

\begin{center}
\begin{tikzpicture}[domain=0:14,scale=0.35]
\draw[very thin,color=white] (-2.5,-1) grid (8.5,5.5);
\node at (-2,5.5) {(d)};

\draw[thin,color=black] (-0.5,-0.5) grid (5.5,5.5);

\draw[thick,color=white] (1,1) -- (1,2);
\draw [thin] (1,2) .. controls (2,2.5) and (2.5,3) .. (3,4);
\draw[thick,color=white] (0,2) -- (-0.5,2);
\draw[thick,color=white] (5,2) -- (5.5,2);
\draw [thin] (5,2) .. controls (4,2.8) and (4,2.8) .. (3,3);

\draw[thick,color=white] (0,4) -- (1,4);
\draw [thin] (0,4) .. controls (1,4.2) and (1,4.2) .. (2,5);

\draw[thick,color=white] (4,1) -- (5,1);
\draw [thin] (4,1) .. controls (3,1.3) and (3,1.3) .. (2,1);

\draw[thick,color=white] (4,0) -- (4,1);
\draw [thin] (4,1) .. controls (5,2) and (4.8,3) .. (5,4);

\foreach \x in {0,...,5}
\foreach \y in {0,...,5}{
\fill (\x,\y) circle (2pt);
}

\end{tikzpicture}
\begin{tikzpicture}[domain=0:14,scale=0.5]
\draw[very thin,color=white] (-4,-2.5) grid (4,2.5);
\node at (-3,2) {(e)};

\path (0: 1cm) coordinate (S1);
\path (60: 1cm) coordinate (S2);
\path (2*60: 1cm) coordinate (S3);
\path (3*60: 1cm) coordinate (S4);
\path (4*60: 1cm) coordinate (S5);
\path (5*60: 1cm) coordinate (S6);

\fill (S1) circle (2pt);
\fill (S2) circle (2pt);
\fill (S3) circle (2pt);
\fill (S4) circle (2pt);
\fill (S5) circle (2pt);
\fill (S6) circle (2pt);

\draw[thin,color=black] (S1) -- (S2);
\draw[thin,color=black] (S2) -- (S3);
\draw[thin,color=black] (S3) -- (S4);
\draw[thin,color=black] (S4) -- (S5);
\draw[thin,color=black] (S5) -- (S6);
\draw[thin,color=black] (S6) -- (S1);

\path (0-15: 2cm) coordinate (S1a);
\path (0: 2cm) coordinate (S1b);
\path (0+15: 2cm) coordinate (S1c);

\path (60-15: 2cm) coordinate (S2a);
\path (60: 2cm) coordinate (S2b);
\path (60+15: 2cm) coordinate (S2c);

\path (2*60-15: 2cm) coordinate (S3a);
\path (2*60: 2cm) coordinate (S3b);
\path (2*60+15: 2cm) coordinate (S3c);

\path (3*60-15: 2cm) coordinate (S4a);
\path (3*60: 2cm) coordinate (S4b);
\path (3*60+15: 2cm) coordinate (S4c);

\path (4*60-15: 2cm) coordinate (S5a);
\path (4*60: 2cm) coordinate (S5b);
\path (4*60+15: 2cm) coordinate (S5c);

\path (5*60-15: 2cm) coordinate (S6a);
\path (5*60: 2cm) coordinate (S6b);
\path (5*60+15: 2cm) coordinate (S6c);

\fill (S1b) circle (2pt);
\fill (S2a) circle (2pt);
\fill (S2b) circle (2pt);
\fill (S2c) circle (2pt);
\fill (S3b) circle (2pt);
\fill (S4b) circle (2pt);
\fill (S6b) circle (2pt);
\fill (S6c) circle (2pt);

\draw[thin,color=black] (S1) -- (S1b);
\draw[thin,color=black] (S2) -- (S2a);
\draw[thin,color=black] (S2) -- (S2b);
\draw[thin,color=black] (S2) -- (S2c);
\draw[thin,color=black] (S3) -- (S3b);
\draw[thin,color=black] (S4) -- (S4b);
\draw[thin,color=black] (S6) -- (S6b);
\draw[thin,color=black] (S6) -- (S6c);

\path (45-10: 3cm) coordinate (S2aa);
\path (45: 3cm) coordinate (S2ab);
\path (45+10: 3cm) coordinate (S2ac);

\fill (S2aa) circle (2pt);
\fill (S2ab) circle (2pt);
\fill (S2ac) circle (2pt);

\draw[thin,color=black] (S2a) -- (S2aa);
\draw[thin,color=black] (S2a) -- (S2ab);
\draw[thin,color=black] (S2a) -- (S2ac);

\path (180: 3cm) coordinate (S4bb);
\fill (S4bb) circle (2pt);
\draw[thin,color=black] (S4b) -- (S4bb);

\path (5*60+15: 3cm) coordinate (S6cb);
\fill (S6cb) circle (2pt);
\draw[thin,color=black] (S6c) -- (S6cb);

\path (30: 4cm) coordinate (S2aaa);
\fill (S2aaa) circle (2pt);
\draw[thin,color=black] (S2aa) -- (S2aaa);
\end{tikzpicture}
\begin{tikzpicture}[domain=0:14,scale=0.5]
\draw[very thin,color=white] (-4,-2.5) grid (4,2.5);
\node at (-3,2) {(f)};

\path (0: 2cm) coordinate (S1);
\path (60: 2cm) coordinate (S2);
\path (2*60: 2cm) coordinate (S3);
\path (3*60: 2cm) coordinate (S4);
\path (4*60: 2cm) coordinate (S5);
\path (5*60: 2cm) coordinate (S6);

\fill (S1) circle (3pt);
\fill (S2) circle (3pt);
\fill (S3) circle (3pt);
\fill (S4) circle (3pt);
\fill (S5) circle (3pt);
\fill (S6) circle (3pt);

\draw[line width=0.07cm,color=black] (S1) -- (S2);
\draw[line width=0.003cm,color=black!80] (S1) -- (S3);
\draw[line width=0.02cm,color=black!100] (S1) -- (S4);
\draw[line width=0.04cm,color=black!100] (S1) -- (S5);
\draw[line width=0.001cm,color=black!70] (S1) -- (S6);
\draw[line width=0.005cm,color=black!80] (S2) -- (S3);
\draw[line width=0.02cm,color=black!90] (S2) -- (S4);
\draw[line width=0.006cm,color=black!90] (S2) -- (S5);
\draw[line width=0.002cm,color=black!80] (S2) -- (S6);
\draw[line width=0.05cm,color=black!100] (S3) -- (S4);
\draw[line width=0.003cm,color=black!80] (S3) -- (S5);
\draw[line width=0.025cm,color=black!100] (S3) -- (S6);
\draw[line width=0.015cm,color=black!80] (S4) -- (S5);
\draw[line width=0.007cm,color=black!90] (S4) -- (S6);
\draw[line width=0.001cm,color=black!70] (S5) -- (S6);

\end{tikzpicture}

\end{center}
\caption{\label{fignetworktypes}(a) A complete, symmetric network on $n=9$ vertices. (b) A random network on $n=12$ vertices with expected mean degree $\hat{d}=4$. (c) A small-world ring on $n=12$ vertices, with $K=4$. (d) A $5 \times 5$ small-world grid. (e) A ring-preferential network on $n=20$ vertices with initial ring size $n_r=6$, initial clique size $K=2$, and attachment degree $d_0=1$. (f) A (symmetric) Gamma network on $n=6$ vertices, with coupling strengths depicted as edge widths. }
\end{figure}

In general, neither infectivity nor susceptibility are constant in the networks described above, the exceptions being complete, symmetric networks, and small-world networks with $p=0$. However, we may modify $\be$ for any connected network to ensure constant infectivity, or constant susceptibility, but not both. To get a {\em constant-infectivity} network, we independently rescale off-diagonal entries in each row of $\be$ to ensure row sums of $1$; and to get a {\em constant-susceptibility} network, we independently rescale off-diagonal entries in each column of $\be$ to ensure column sums of $1$. These procedures can be applied also to individual-based networks. Note that for constant-infectivity and constant-susceptibility networks, $\be$ is no longer, in general, a symmetric matrix. As we will see below, many qualitative conclusions hold for both the default networks, and their constant-infectivity or constant-susceptibility versions, and are hence associated with the network topology, rather than consequent variations in infectivity or susceptibility. Some conclusions, however, are sensitive to whether we allow variations in susceptibility or not. 

\subsection{Simulations}

All simulations to follow are carried out in {\tt C/C++}, with all code and scripts available on GitHub \citep{muradEpigithub}. For simplicity we fix $\wR_0$, or expected $\wR_0$ in the case of Gamma networks, to be $2$ in all simulations unless stated otherwise. Other parameters are chosen to illustrate qualitative behaviours, and allow easy comparison of the effects of variations in network architecture, network parameters or prior immunity on outbreak size. Simulation methodologies and parameter choices are discussed further in Appendix~\ref{appsim}.

\section{Results}

\subsection{Measures of population protection}
\label{secprotect}

We need some natural measure which will allow us to compare population vulnerability across spatial models with different architectures. The classical effective reproduction number, $\wR_t$, defined in (\ref{classicRt}), and the next-generation effective reproduction number $\mathcal{R}_t$, defined in (\ref{nextgenRt}), provide only limited information. For example, $\wR_t$ can drop below $1$ even as an outbreak is surging; and $\mathcal{R}_t$ can remain close to $\mathcal{R}_0$, even as outbreaks become unlikely and/or small, see Figure~\ref{figRt} below. As an alternative to $\wR_t$ and $\mathcal{R}_t$, we could approximate the vector of outbreak probabilities using the branching process approximation (see Appendix~\ref{appbranching}); but these approximations are only valid when subpopulations are large, and suffer the drawback that a high outbreak probability may be associated with a small expected outbreak size, or the reverse. 

With these considerations in mind, we choose {\em mean outbreak size}, as a measure of population vulnerability which is easy to interpret inutitively and to compare across models. It can in some simple cases be estimated analytically; but for the most part we compute mean outbreak size numerically, in which case we denote the number of stochastic simulations used to compute each mean value by $n_{sim}$. 

In most cases, we present mean outbreak size as a percentage of the total population, but it is sometimes instructive to normalise it against the expected outbreak size in the homogeneous case. For this purpose, we define $E(s,\wR_0)$ to be the outbreak size as a fraction of the total population in a homogeneous, deterministic SIR model with initial susceptible fraction $s$, and basic reproduction number $\wR_0$, namely, 
\begin{equation}
  \label{lambertmain}
E(s, \wR_0) = s+\wR_0^{-1}W(-s\wR_0e^{-s\wR_0})\,,
\end{equation}
where $W(\cdot)$ is the Lambert $W$ function. In the stochastic setting,
\begin{equation}
  \label{lambertmain1}
E^*(s, \wR_0) = \max\left\{1-\frac{1}{s\wR_0},0\right\} \times \left(s+\wR_0^{-1}W(-s\wR_0e^{-s\wR_0})\right)\,
\end{equation}
is then the {\em unconditional} expected outbreak size (as a fraction of the total population) given an introduction into a large, homogeneous population with susceptible fraction $s$.

\begin{figure}[htp!]

  \begin{center}
    \underline{Tracking measures of population vulnerability during an outbreak}
    
  \includegraphics[scale=0.9]{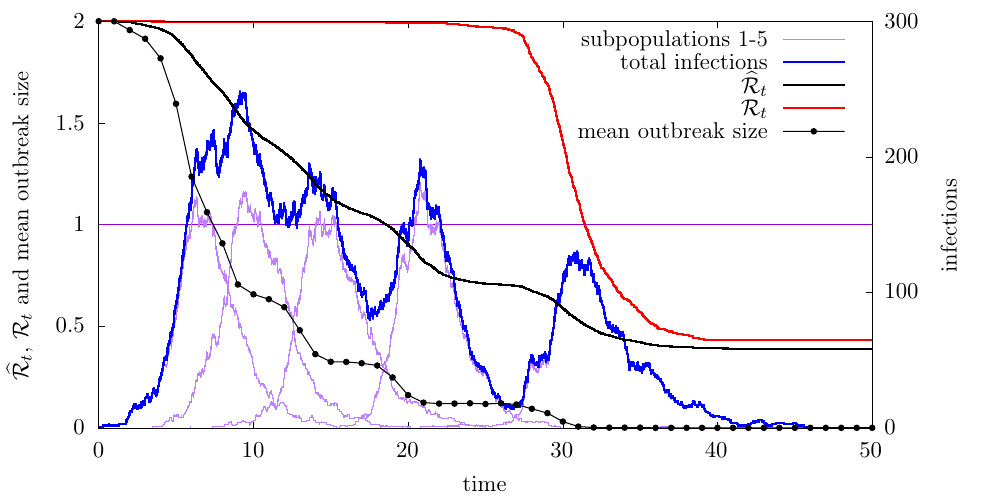}
\end{center}
\caption{\label{figRt}An outbreak hitting all subpopulations in a complete, symmetric network with $n=5$, $N_c=1000$, $\widehat{\mathcal{R}}_0 = 2$, and $\Theta N_c = 2$. During the outbreak we track the classical effective reproduction number $\wR_t$ (red), the next generation effective reproduction number $\mathcal{R}_t$ (black) and the mean secondary outbreak size ($n_{sim} = 50000$) scaled to have initial value $2$ for easy comparison with $\wR_t$ and $\mathcal{R}_t$.}
\end{figure}

In Figure~\ref{figRt}, we illustrate how $\wR_t$, $\mathcal{R}_t$ and mean outbreak size, computed numerically from $n_{sim} = 50000$ random introductions, behave during an outbreak in a complete, symmetric network with five weakly-coupled subpopulations. Note, first, that outbreak trajectories in spatial models can look very different from the standard bell-shaped SIR trajectories of homogeneous models. Recall that, as the network is a constant-susceptibility network, by (\ref{RtclassA}), $\wR_t$ is proportional to the total susceptible fraction: in this example $\wR_t$ crosses below $1$ while infections are rising. On the other hand, $\mathcal{R}_t$ only decreases noticeably once the final subpopulation is hit, after which it declines rapidly. The mean (secondary) outbreak size, given in arbitrary units, drops most sharply each time a new subpopulation is hit, effectively reaching zero when $\mathcal{R}_t$ crosses below $1$. 

Figure~\ref{figRt} demonstrates the challenges of defining a herd immunity threshold in stochastic models with finite subpopulations. In this context, we may adopt a practical definition based on mean outbreak sizes. Fix $h \ll 1$. In a given fully susceptible population, let us define $E$ to be the expected outbreak size given a random introduction. Now, given partial immunisation of the population, we may define the population to have achieved herd immunity at threshold $h$ if the expected outbreak size is no more than $hE$.

We now examine how mean outbreak sizes depend on the network architecture, subpopulation size, coupling strength, and prior immunity, both infection-acquired and random. All computational details are as described in Appendix~\ref{appsim}.

\subsection{Outbreak sizes as a function of leak and subpopulation size}

We start with the most intuitively obvious claim: for sufficiently small leak, and sufficiently large subpopulations, outbreak sizes increase with leak. We can demonstrate this by choosing a network of any fixed architecture, increasing the mean leak $\Theta$, and examining the mean outbreak size. 

\begin{figure}[h!]

  \begin{center}
    \underline{Normalised outbreak size versus leak for various network architectures}

  \includegraphics[scale=0.27]{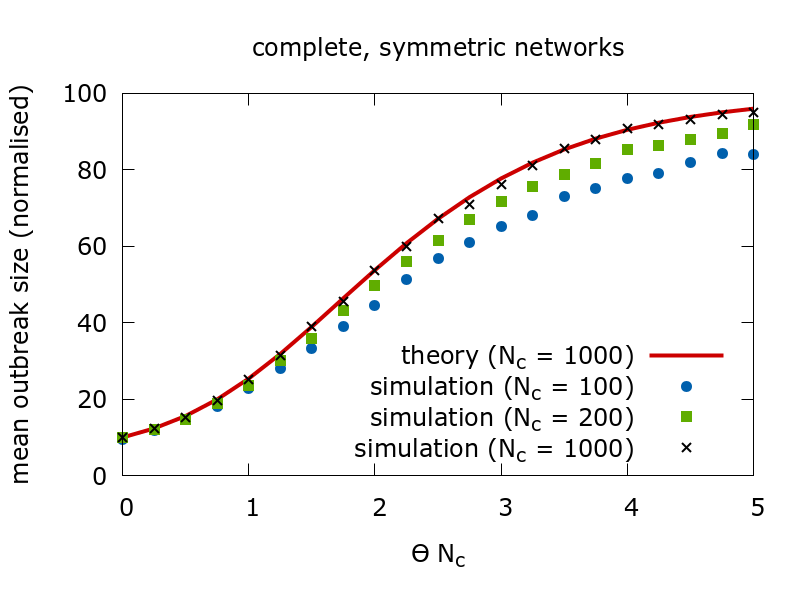}
  \includegraphics[scale=0.27]{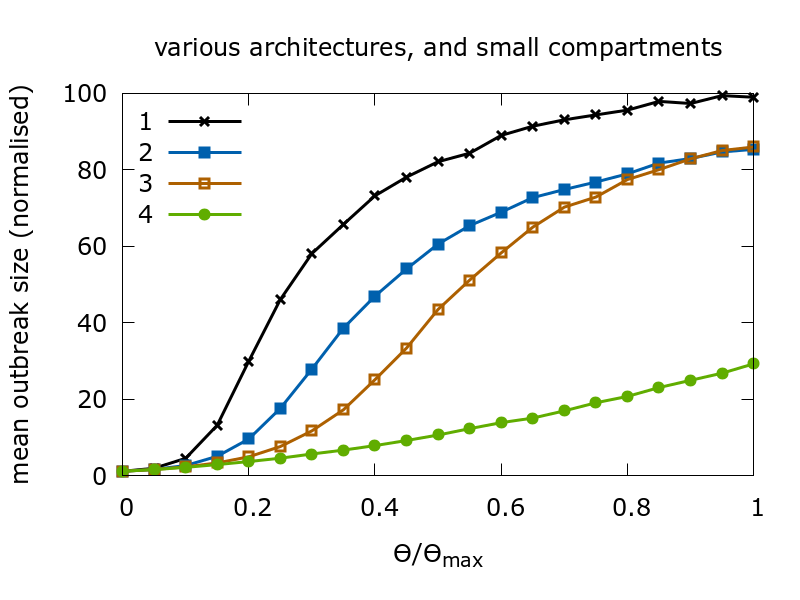}
\end{center}

\caption{\label{fig_leak1}Mean normalised outbreak sizes as we vary the leak. The normalisation is against the unconditional expected outbreak size in the homogeneous case, as in \eqref{lambertmain1}. In each case, $n_{sim} = 20000$. {\em Left.} Complete, symmetric networks with $n=10$, and $N_c = 100$, $N_c=200$ or $N_c=1000$. The analytical approximation is based on results in Appendix~\ref{appcomplete}, and matches simulations well in the case $N_c=1000$. For $N_c = 100$ and $N_c = 200$, the analytical estimates (not shown) are close to those for $N_c = 1000$ and overestimate outbreak sizes. {\em Right.} Various networks with $n=100$ and $N_c=10$. (1) a complete, symmetric network; (2) a random network with expected mean degree $\hat{d}=4$; (3) a $10 \times 10$ grid with no rewiring; and (4) a ring with clique size $4$ and no rewiring.}
\end{figure}

In Figure~\ref{fig_leak1} we demonstrate the dependence of outbreak size on mean leak across networks with different architectures. In complete, symmetric networks (Figure~\ref{fig_leak1}, left), we can compare the simulations with analytical estimates derived using the branching process approximation in Appendix~\ref{appcomplete}. We see a good match between simulation and analysis when $N_c$ is large.

At $\Theta = \Theta_{max}$ a complete, symmetric $n$-subpopulation network is, formally, a one-subpopulation network, and expected outbreak size should exactly match the homogeneous case which, for a sufficiently large total population, will be well-approximated by \eqref{lambertmain1}. However, for other architectures, we find that mean outbreak sizes can remain below the theoretical maximum for all $\Theta \in [0,\Theta_{max}]$, see Figure~\ref{fig_leak1} (right). We may explain this by noting that even when $\Theta = \Theta_{max}$, stochastic extinction means that outbreaks can remain localised. This is particularly pronounced in the ring lattice which has a large mean path length. 

Percolation theory (\cite{Kenahpercolation,Millerpercolation,Moorepercolation} for example) tells us that in large, individual-based networks, we expect some threshold value of transmissibility, dependent on the network architecture, at which outbreak sizes grow rapidly to become comparable with the population size. A similar {\em invasion threshold} can be observed in heterogeneous metapopulation models \citep{Colizza2007} as the diffusion rate increases. In our models, finite subpopulations replace individuals, the number of compartments is finite, and the leak plays a role analogous to transmissibility or diffusion. While we do not expect the sharp phase transitions of percolation theory, by analogous arguments, we nevertheless might expect a sigmoidal response of outbreak size to increasing leak. Indeed, this is observed across network topologies, and the sigmoidal response becomes sharper as subpopulations grow larger, see Figure~\ref{figsigmoidal} in Appendix~\ref{secgraphs}. 

Additionally, Figure~\ref{figsigmoidal} in Appendix~\ref{secgraphs} highlights the fundamental effect of the compartment size $N_c$ on mean outbreak size, with smaller compartments associated with higher probabilities of stochastic extinction, hence smaller outbreaks at any fixed level of the leak $\Theta$.

In a real-world context, we might interpret the sigmoidal response of outbreak size to leak in terms of {\em non-pharmaceutical interventions}, for example restrictions on mobility which reduce contact between different subpopulations. In a metapopulation composed of small subpopulations, we expect an approximately linear decrease in outbreak size as the mean leak, namely, the fraction of infections occurring between subpopulations, decreases; with larger subpopulations we expect a more clearly nonlinear response, with a sharp change in outbreak size near some critical level of inter-subpopulation transmission. 

We make a couple of additional remarks. Whenever $\wR_0 > 1$, and for any given strongly connected network and any sufficiently small leak, expected outbreak sizes approach the value given by \eqref{lambertmain1} as $N_c \to \infty$. Intuitively, even if the leak is small, given a sufficiently large outbreak in one subpopulation, the probability of contagion to neighbouring subpopulations approaches $1$. This claim is straightforward to formalise using the techniques in Section~\ref{seccontagion1}. Secondly, for networks whose architecture and coupling strength imply highly variable infectivity, expected outbreak size can exceed the value given by \eqref{lambertmain1} for a homogeneous network with the same value of $\wR_t$; see Figure~\ref{fig_vaxreversed} in Appendix~\ref{secgraphs}, for example. We can understand this intuitively by considering that even in states such that $\wR_t \approx 1$, $\mathcal{R}_t$ may be greater than $1$, with infection able to spread in some parts of the network, leading to substantial mean outbreak size, see also \cite[Chapter 12]{AndersonMay1991}.

\subsection{The effects of network architecture on mean outbreak size}
Network achitecture, independently of coupling strength, has a strong effect on expected outbreak size. Broadly, as we increase the fraction of ``non-local'' links in a network, local herd immunity effects become less important, and larger outbreaks occur on average, even if the mean leak is held constant. 

\begin{figure}[h]
  \begin{center}
    \underline{Increasingly de-localised outbreak patterns in small-world grids}
    
  \includegraphics[scale=0.4]{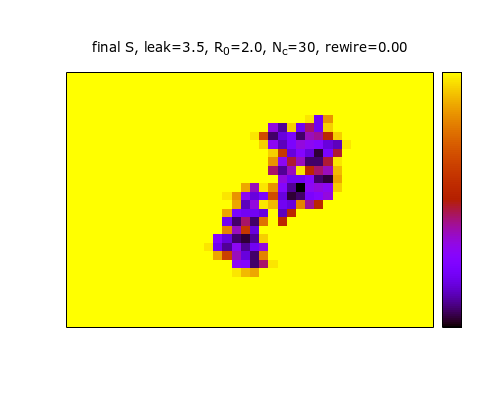}
  \includegraphics[scale=0.4]{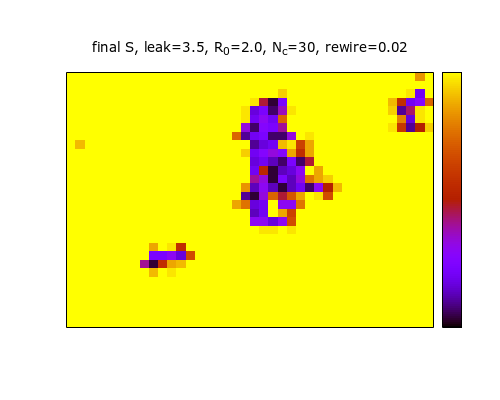}\\\vspace{-0.5cm}
  \includegraphics[scale=0.4]{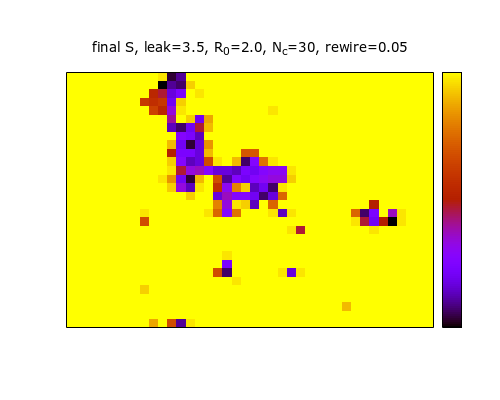}
  \includegraphics[scale=0.4]{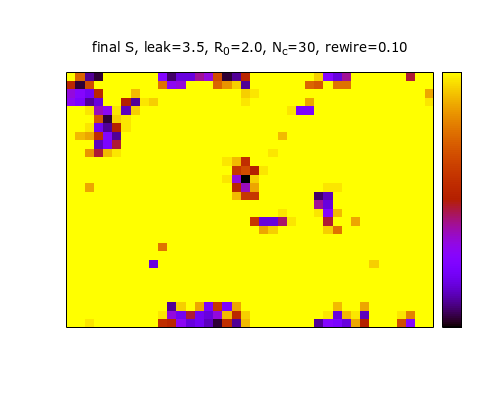}\vspace{-0.5cm}
\end{center}

\caption{\label{fig_swgrid}Outbreak patterns in small-world grids as we increase the rewiring probability. In each case we have a $40 \times 30$ grid with $N_c = 30$ and $\Theta N_c=3.5$. Read lexicographically from top left, the rewiring probability increases from $0\%$, to $2\%$, to $5\%$, to $10\%$. The final population state at the end of typical outbreaks of approximately the same size are shown, with yellow indicating a fully susceptible population and darker colours indicating where infection has spread. Each plot is associated with a corresponding video on GitHub \citep{muradEpigithub}.}
\end{figure}

We can most easily visualise this claim in small-world grids with increasing rewiring. We see in Figure~\ref{fig_swgrid} how even a small positive rewiring probability $p$, leading to a few long-range links, tends to create increasingly de-localised outbreaks.

\begin{figure}[htp!]
  \begin{center}
\underline{Outbreak size as we vary network architecture (default networks)}
    
  \includegraphics[scale=0.4]{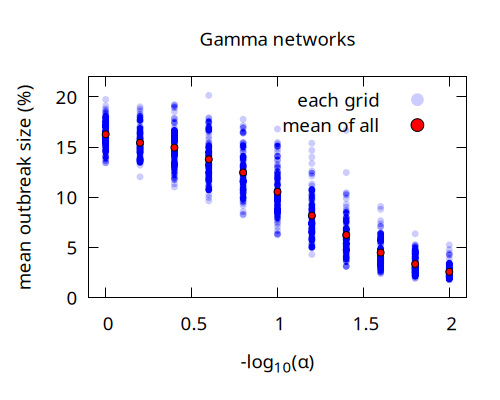}
  \includegraphics[scale=0.4]{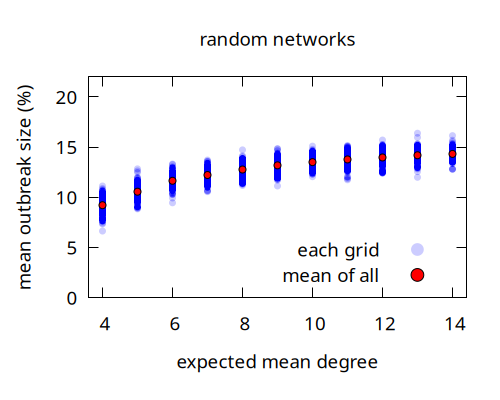}
  \includegraphics[scale=0.4]{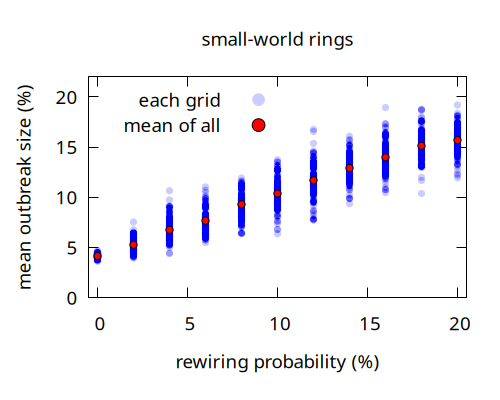}
  \includegraphics[scale=0.4]{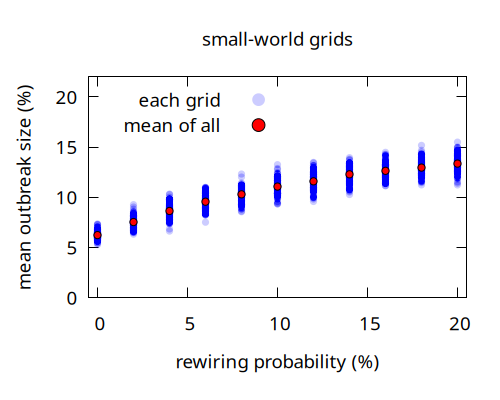}
  
\end{center}

\caption{\label{figarchitecture}For each parameter value, random introductions to $100$ randomly chosen networks were simulated $n_{sim}=1000$ times each to determine the mean outbreak size (blue dots), with the global mean shown as a red dot. {\em Top left.} (Asymmetric) Gamma networks with $n=30, N_c=50$, $\Theta N_c = 2.5$ and variable shape parameter $\alpha$.
  {\em Top right.} Random networks with $n=100, N_c=50$, $\Theta N_c = 2.5$, and variable expected mean degree. 
{\em Bottom left.} Small-world rings with $n=100$, $K=4$, $N_c=50$, $\Theta N_c = 4$, and variable rewiring probability. {\em Bottom right.} A small-world $19 \times 13$ grid with $N_c = 30$, $\Theta N_c = 3.5$, and variable rewiring probability.}

\end{figure}

The way that network architecture quantitatively affects outbreak size is illustrated for four different architectures in Figure~\ref{figarchitecture}. In each case, we fix the class of networks and the mean leak, but alter some parameter which effectively changes the wiring of the network. Parameters are chosen so that maximal outbreak sizes are similar in each case.

In a Gamma network, if we maintain the expected mean leak $\Theta$, but increase the variance of the leaks $\varepsilon_{ij}$, this has an effect analogous to increasing mean path length in the network. In Figure~\ref{figarchitecture}, top left, we decrease the shape parameter $\alpha$ of the distribution from which we choose $\varepsilon_{ij}$, hence increasing its variance $\Theta^2/(\alpha(n-1)^2)$. For sufficiently large $\alpha$, the network behaves similarly to a complete symmetric network; however, as $\alpha$ decreases, outbreaks become more localised, hence smaller. (Although not shown, the pattern is similar for symmetric Gamma networks.) 

In random networks, increasing the expected mean degree, while maintaining the mean leak, increases outbreak sizes (Figure~\ref{figarchitecture}, top right). In small-world networks (Figure~\ref{figarchitecture}, bottom left and right) outbreak sizes increase as the rewiring probability increases and outbreaks become less localised, as illustrated for small-world grids in Figure~\ref{fig_swgrid}.

We see the same effects as in Figure~\ref{figarchitecture} if, instead of the default networks, we use constant-infectivity or constant-susceptibility networks, see Figures~\ref{figarchitecture1}~and~\ref{figarchitecture2} in Appendix~\ref{secgraphs}. 

Network architecture affects not only expected outbreak size, but also other qualitative features of the distribution of outbreak sizes. As an example, the effect of increasing the rewiring probability $p$ on the distribution of outbreak sizes in small-world grids is shown in Figure~\ref{figsizehist} in Appendix~\ref{secgraphs}. Increasing $p$ results in a more clearly bimodal distribution of outbreak sizes. A strongly bimodal outbreak size distribution allows us easily to separate minor and major outbreaks, and is associated with a larger fraction of major outbreaks leading to herd immunity at any reasonable threshold (see Section~\ref{secprotect}).

\subsection{Prior random immunity is often more protective in spatial models}
\label{seccompvhom}
We define a {\em random immunisation} to be an immunisation which occurs in subpopulation $i$ with probability proportional to the fraction of the total susceptible population in subpopulation $i$, namely, $S_i/S_{tot}$; and immunising a certain fraction of the population randomly will mean carrying out a sequence of random immunisations until the desired immune fraction is achieved. 

Our next observation is that random immunity frequently has a greater protective effect in a spatial model than in its homogeneous counterpart with the same value of $\wR_0$. This conclusion holds across a variety of model architectures.

\begin{figure}[h]
\begin{center}
\underline{Outbreak sizes as a function of prior random immunity}

  \includegraphics[scale=0.37]{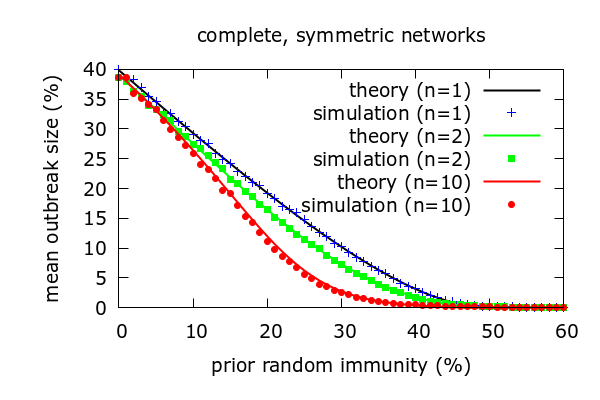}
  \includegraphics[scale=0.37]{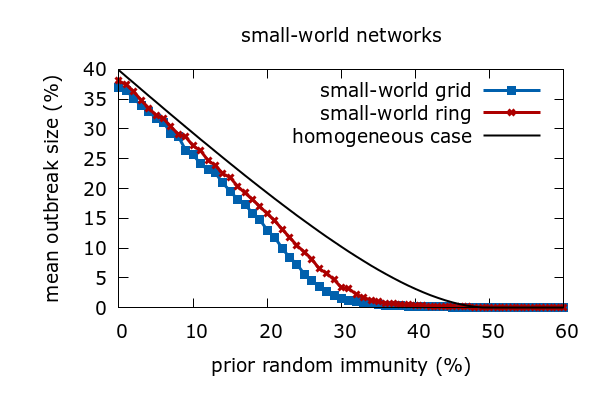}
  \includegraphics[scale=0.37]{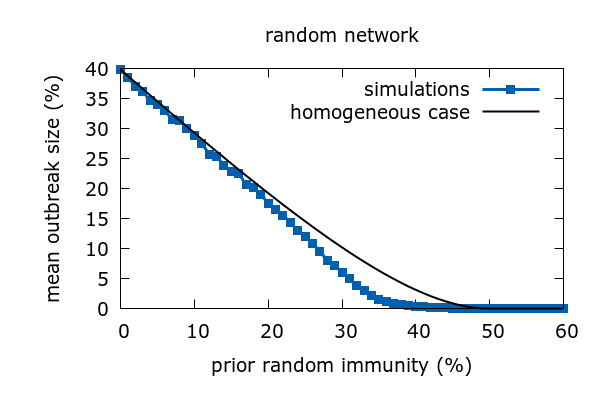}  
  \includegraphics[scale=0.37]{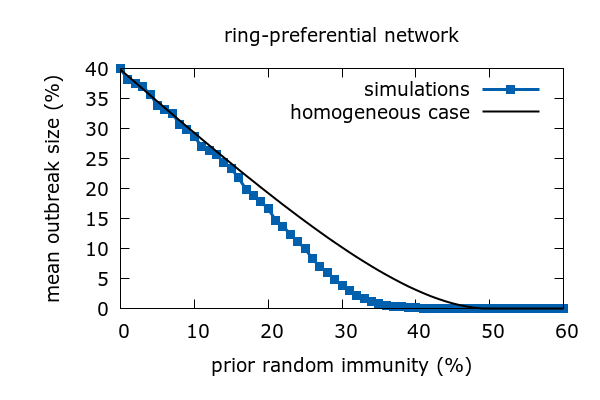}  
\end{center}

\caption{\label{fig_vax10c}Mean outbreak sizes as a function of prior random immunity in various networks. {\em Top left.} Complete, symmetric networks with total population $nN_c=10000$, and $n=1$ ($\Theta$ undefined), $n=2$ ($\Theta N_c=4$) and $n=10$ ($\Theta N_c = 6$). {\em Top right.} Two small-world grids with $n=100$ and $N_c=100$: a $10 \times 10$ grid with $\Theta N_c=15$ and rewiring probability of $5\%$, and a ring with clique size $4$, $\Theta N_c=25$, and rewiring probability of $10\%$. {\em Bottom left.} A random network with $n=25$, $N_c=1000$, $\Theta N_c = 20$, and expected mean degree $\hat{d} = 5$. {\em Bottom right.} A ring-preferential network with $n=50$, $N_c = 2000$, $\Theta N_c = 20$, initial ring size 6, clique size $2$, and attachment degree $1$.}
\end{figure}

The claim is illustrated for complete, symmetric networks in Figure~\ref{fig_vax10c} (top left). As we increase the level of random immunity, the mean outbreak size drops more rapidly in the spatially stratified case than in the homogeneous case, with the effect being stronger when we have more subpopulations. The analytical approximations derived for complete, symmetric networks with large subpopulations and weak coupling in Appendix~\ref{appcomplete} match well with simulations. In Figure~\ref{fig_vax10c} (top right), we see the same effect for small-world grids. This holds even though the leak is chosen to be large enough so that, in the absence of prior immunity, mean outbreak sizes are similar to the homogeneous case. In Figure~\ref{fig_vax10c} (bottom left and bottom right), we see that the effect occurs also in random networks and ring-preferential networks with sufficiently large subpopulations and weak coupling between subpopulations.

We may interpret the results of Figure~\ref{fig_vax10c} in terms of herd immunity, as discussed in Section~\ref{secprotect}. We see that dividing a homogeneous population into a network of subpopulations can significantly reduce the herd immunity threshold under random immunisation without necessarily reducing mean outbreak sizes for a fully susceptible population, and that this can occur for a variety of choices of network topology, and for arbitrary values of the threshold $h$.

To understand why, in spatial models, the protective effect of relatively small amounts of prior random immunity can be greater than in the homogeneous case, we note that for any network architecture adding random immunity decreases both: (i) the local expected outbreak size in any given subpopulation; and (ii) the probability of contagion between subpopulations. The latter claim follows, in the large population, weak coupling, limit, from Equations~\ref{eqW}~and~\ref{eqalpha1}. Recall that $\alpha_{ij}(m) = \alpha_{ij}(1)^m$ is the approximate probability of no contagion from subpopulation $i$ to subpopulation $j$ given $m$ infections in subpopulation $i$. Decreasing the susceptible fraction in subpopulation $j$ decreases $\wW_{ij}$ thus increasing $\alpha_{ij}(1)$; meanwhile, decreasing the susceptible fraction in subpopulation $i$ decreases the expected size of any outbreak in subpopulation $i$, namely the exponent $m$. Both effects lead to increases in $\alpha_{ij}(m)$. 

We observe, however, that the claim that random immunity provides greater benefits in spatial models than in homogeneous models can fail for some architectures and sufficiently strong coupling. To understand why, observe that prior random immunity leading to $\wR_t \approx 1$ would imply small expected outbreak sizes in the homogeneous case but not necessarily in heterogeneous networks where we may still have have $\mathcal{R}_t$ significantly above $1$. In Figure~\ref{fig_vaxreversed} in Appendix~\ref{secgraphs} we present examples where the effect demonstrated in Figure~\ref{fig_vax10c} fails to hold, and prior random immunity provides similar or greater protection in a homogeneous population than in a networked population. See \citep[Chapter 12]{AndersonMay1991}, and \citep{PastorSantorras2002} for further examples of this latter effect.

\subsection{Outbreak sizes following infection-acquired and random immunity}

We next show that for a variety of network architectures infection-acquired immunity provides less population protection than the same level of random immunity. On the other hand, if we construct networks with sufficient heterogeneity in susceptibility, then this result can be reversed. These claims are illustrated in Figure~\ref{fig_protectmain}, while the details are explored further in a number of further simulations presented in Appendix~\ref{secgraphs} and discussed below.

\begin{figure}[htp!]
\begin{center}
  \underline{Outbreak size versus prior immunity in default networks}
    
  \includegraphics[scale=0.35]{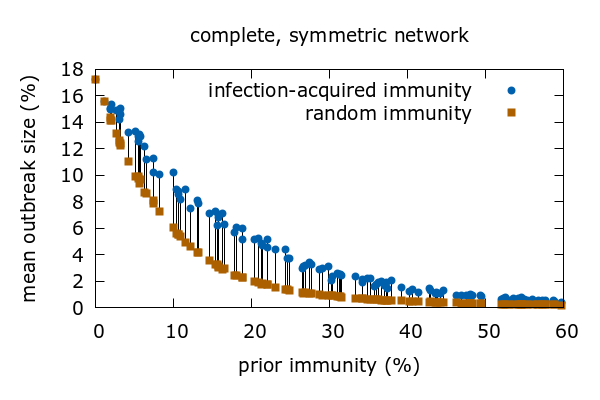}
  \includegraphics[scale=0.35]{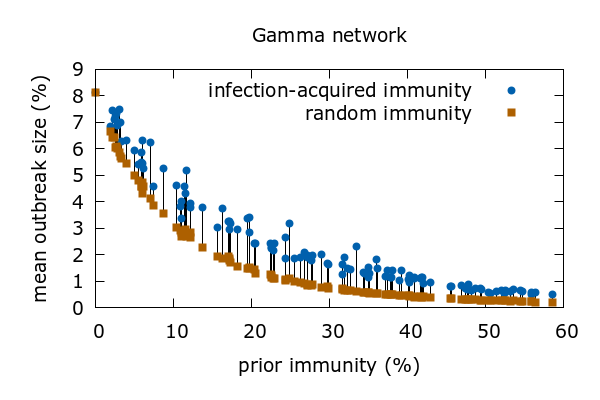}
  \includegraphics[scale=0.35]{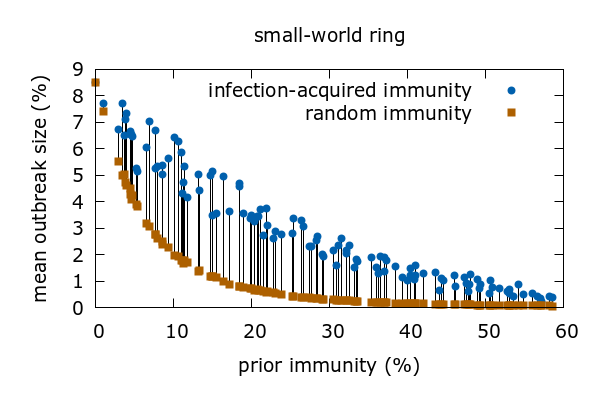}
  \includegraphics[scale=0.35]{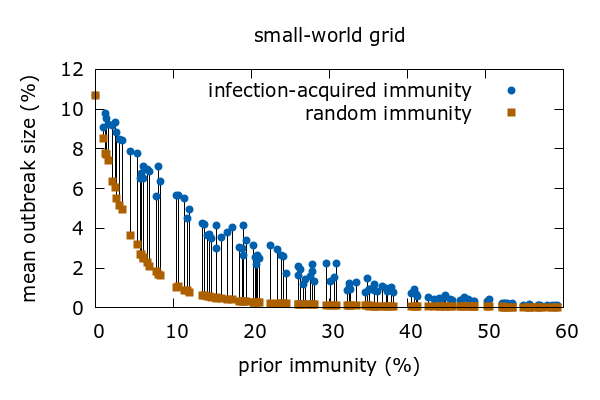}
  \includegraphics[scale=0.35]{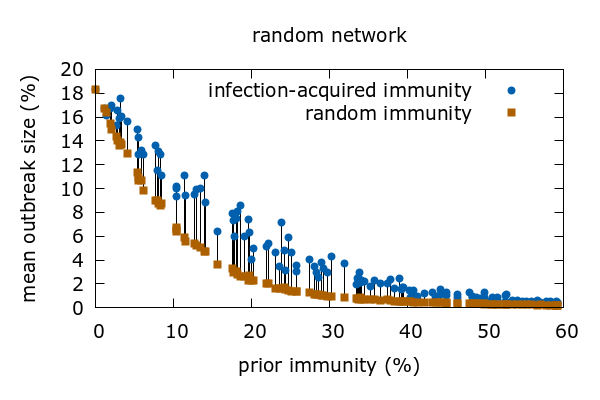}
  \includegraphics[scale=0.35]{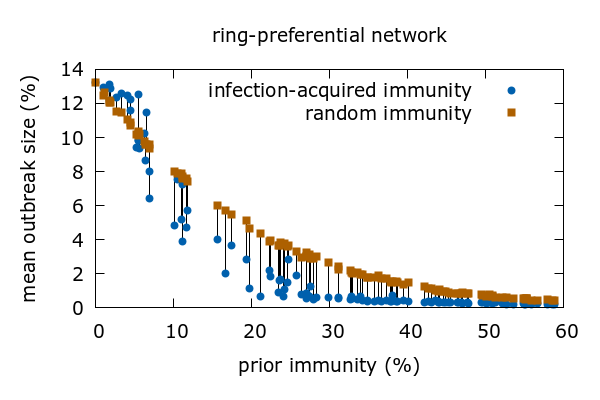}
\end{center}

  \caption{\label{fig_protectmain}Mean outbreak sizes following an outbreak, or following addition of the same amount of random immunity. {\em Top left.} A complete, symmetric network with $n=30, N_c=50$ and $\Theta N_c=2.5$.
    {\em Top right.} An (asymmetric) Gamma network with $n=30$, $N_c = 50$, $\Theta N_c = 2.5$ and shape parameter $0.1$.
    {\em Middle left.} A ring small-world network with $n=100$, $N_c=50$, $\Theta N_c = 4$, clique size $4$ and rewiring probability $10\%$.
    {\em Middle right.} A regular $19 \times 13$ grid with $N_c=50$, $\Theta N_c = 3.5$, and rewiring probability $5\%$.
  {\em Bottom left.} A random network with $n=30$, $N_c=50$, $\Theta N_c = 3.0$, and expected mean degree $\hat{d} = 8$. 
  {\em Bottom right.} A ring-preferential network with $N_c = 5$, $n=200$, $\Theta N_c=2.5$, initial ring size 10, clique size $4$, and attachment degree $1$.
  }
\end{figure}

In order to compare the effects of random immunity versus infection-acquired immunity, we follow a procedure close to that in \cite{Ferrari}, where individual-based networks were studied. In brief, we initiate primary outbreaks, let these run their course, and then initiate secondary outbreaks with a new introduction; we compare the sizes of these secondary outbreaks to outbreaks in the same network with the same amount of prior {\em random} immunity as resulted from the primary outbreak. Further details are in Appendix~\ref{appsim}.

The first conclusion we may infer from Figure~\ref{fig_protectmain} concerns our claim that in spatial models outbreaks often end without herd immunity. To see this, note that each (blue) circle corresponds to a primary outbreak and a set of secondary outbreaks. In all the architectures, we find that mean secondary outbreak size can be considerable. To quantify how often this occurs, we set some threshold $h$ as in Section~\ref{secprotect}, and define a ``large'' outbreak as an outbreak of size at least $h$ times the expected outbreak size for the fully susceptible population. We then initiate outbreaks and for each large outbreak calculate the mean secondary outbreak size. If the mean secondary outbreak is large, we say that the network has failed to reach herd immunity at the threshold $h$. Some simulation results are shown in Table~\ref{table_herd}. We see, for example, that in a small-world grid, with parameters as in Figure~\ref{fig_protectmain}, only about 50\% of large outbreaks lead to herd immunity at $h=0.1$.

\begin{table}[h]\centering
\begin{center}
  \underline{The fraction of outbreaks leading to herd immunity}
 \end{center}
 \ra{1.3}
\begin{tabular}{lcc}\toprule
network architecture & herd immunity achieved (\%) at $h=0.1$\\ \midrule
complete, symmetric network &  59\%\\
gamma network & 18\%\\
small-world ring & 33\%\\
small-world grid & 50\%\\
random network & 67\%\\
ring-preferential network & 88\%\\
\bottomrule
\end{tabular}

\caption{\label{table_herd}The percentage of outbreaks which lead to herd immunity for a variety of network architectures, at threshold $h=0.1$, and with all parameters as in Figure~\ref{fig_protectmain}. Further details are in the text and in Appendix~\ref{appsim}.}
\end{table}

The second, and main, conclusion that we can draw from Figure~\ref{fig_protectmain} is that, across a range of network architectures, prior random immunity can confer greater protection than infection-acquired immunity. However, sometimes (e.g., Figure~\ref{fig_protectmain}, bottom right) the pattern can be reversed, with infection-acquired immunity being more protective than random immunity. 

We may understand the results depicted in Figure~\ref{fig_protectmain} as follows. In many cases, the reduced probability of contagion arising from a given amount of prior random immunity (see Section~\ref{seccompvhom}), seems to outweigh in importance the ``network frailty'' effects described in \cite{Ferrari}, namely the reduced contagion associated with a decrease in connectivity and increase in mean path length from prior immunity following infection spreading through the network. However, in some situations, network frailty effects are more important, as in Figure~\ref{fig_protectmain}, bottom right. In order to better understand when this occurs, a number of further simulations were carried out. 

The role of subpopulation size is seen in Figures~\ref{fig_protectrandindividual}~and~\ref{figringpref} in Appendix~\ref{secgraphs}. In simulations of individual-based {\em random} networks, we see no advantage in random immunity versus infection-acquired immunity; but as subpopulation size increases, random immunity becomes more protective (see Figure~\ref{fig_protectrandindividual}). In simulations of {\em ring-preferential} networks, as we increase subpopulation size we observe a switch from infection-acquired immunity being more protective to random immunity being more protective (see Figure~\ref{figringpref}). While important, a small subpopulation size is not sufficient for network frailty effects to dominate: in fairly regular individual-based networks, including small-world grids with significant rewiring, random immunity remains more protective than infection-acquired immunity, see Figure~\ref{fig_protectgridindividual} in Appendix~\ref{secgraphs}.

There appears to be a close connection between variability in susceptibility and network frailty. Even in individual-based networks with highly variable degree, such as ring-preferential networks, if we impose constant susceptibility, then random immunity becomes more protective, see Figure~\ref{fig_protectrsfindividual} in Appendix~\ref{secgraphs}. That variations in susceptibility are key to network frailty is consistent with the findings of \cite{GomesJTB}, and an interesting avenue for further exploration. For completeness, and to highlight this conclusion, constant-infectivity and constant-susceptibility versions of the simulations in Figure~\ref{fig_protectmain} are presented in Figures~\ref{fig_protectmainalt}~and~\ref{fig_protectmainalt1} in Appendix~\ref{secgraphs}.

\section{Discussion and conclusions}

During infectious disease outbreaks, outbreak dynamics are heavily dependent on the network structure of the host population \citep{PastorSantorras2015}. This network structure is affected by non-pharmaceutical interventions such as restrictions on mobility; and in turn changes in the network structure can modify the impacts of pharmaceutical interventions such as vaccination. Thus, understanding the potential effects of network structure on transmission is crucial for understanding outbreak dynamics in networked populations, and for planning outbreak mitigation strategies.

We began by showing that, in networked populations, expected outbreak sizes tend to be lower than in homogeneous populations with the same basic epidemiological parameters, although there can be exceptions to this rule, see Figure~\ref{fig_vaxreversed} (right) and see also the discussion in \cite{Leibenzon2024}. Consistent with analytical results for weighted individual-based models \citep{Goltsev2012}, we expect smaller outbreaks in more poorly connected networks, as disease tends to remain more localised. Even with substantial transmission between subpopulations, spatial effects can reduce the mean outbreak size relative to that expected in a homogeneous population. This effect is a consequence of stochastic extinction, and is most pronounced when subpopulations are small and networks are poorly connected.

Apart from mean outbreak sizes, it is important also to consider outbreak size {\em distributions} in network models (see \cite{Neal2003} for a rare example of an exact result). We find that heterogeneity can result in ``flatter'' outbreak size distributions than in the homogeneous case. The most important practical consequence is that a significant fraction of outbreaks in heterogeneous populations may end without the population achieving herd immunity by any reasonable measure. While the distribution of outbreak sizes is strongly dependent on the network details, these results highlight the importance of not rushing to conclude that herd immunity is reached when outbreaks end (see Table~\ref{table_herd} and Figure~\ref{figsizehist} in Appendix~\ref{secgraphs}).

If we fix a network architecture, but increase transmission between subpopulations, outbreak sizes generally increase. However, the response of the mean outbreak size to the extent of coupling between subpopulations can be more or less nonlinear, implying that reductions in mobility may sometimes be ineffective below some threshold level, and become highly effective beyond this level. This conclusion holds across different network architectures.

In contrast to conclusions in \citep[Chapter 12]{AndersonMay1991}, and in \citep{PastorSantorras2002}, we found that the protective effects of random immunity can be greater in spatially structured populations than in homogeneous populations. This is especially true in weakly coupled populations across different architectures: a relatively small amount of random immunity not only reduces the mean outbreak size in each subpopulation, but also reduces the probability of contagion between subpopulations. Thus, reductions in mobility which may not be sufficient in themselves to lower outbreak sizes substantially could nevertheless enhance the protective effect of a limited amount of vaccination. This could be particularly relevant for a vaccine that is only partially effective at reducing transmission, or in the context of the stage of an outbreak soon after a vaccine becomes available. 

Consistent with the recent work of \cite{Hiraoka2025}, but in contrast to findings in  \citep{BrittonCOVID,Montalban,GomesJTB}, we found that in many network models random immunity is more protective than infection-acquired immunity. The expected outbreak size is determined by both the likelihood and size of local outbreaks {\em and} the likelihood of contagion. In fairly regular networks, random immunity provides considerably more protection than infection-acquired immunity. On the other hand, in networks with small subpopulations and heavy-tailed degree distributions, network frailty effects can become important, reversing this effect. Our simulations also demonstrate an intriguing asymmetry in the roles of susceptibility and infectivity: they suggest that it is large variability in susceptibility, rather than infectivity, which is associated with network frailty. This merits further exploration.

Turning to practical conclusions, we should never rush to infer herd immunity when pathogen extinction occurs at relatively low levels of infection -- as happened in some locations during the COVID-19 pandemic -- especially in geographies that are relatively disconnected, either naturally, or as a result of restrictions on mobility. Reduced outbreak sizes can be, at least partly, a consequence of disease localisation; hence, they do not automatically imply that a population is well-protected after an outbreak. A second practical conclusion is that, during any specific outbreak, assertions that infection-induced immunity is more protective than (random) vaccination, or vice versa, should be tested carefully using a detailed model for the specific population and pathogen under consideration.

The contrast between some of our conclusions on random immunity in heterogeneous populations, and previous conclusions in the literature, can be traced to the classes of models which have received most attention. For example, metapopulation models have received less attention than unweighted individual-based models. And there has been a particular emphasis in the stochastic literature on scale-free architectures where random immunisation is ineffective, and network-frailty effects are particularly important. It is notable, however, that outbreak dynamics in networks with contact structures informed by real-world data \citep{Meyers2005,Kitsak2010} can differ substantially from those in scale-free networks constructed via random procedures.

As in any modelling study, we made a number of simplifying assumptions. For example, we considered only a few idealised network structures rather than attempting to model any specific real-world host population. We assumed that all subpopulations were identical, following SIR model dynamics, and ignored explicit diffusion of individuals between subpopulations. In low-prevalence settings early in the COVID-19 pandemic, a key question was whether vaccination would be able to prevent outbreaks \citep{sachak2021risk}. Motivated by such questions, and previous modelling studies on herd immunity, we made the simplifying assumption that random immunity against infection characterises vaccination, and examined the effects of random vaccination on outbreak sizes.

In reality, epidemiological dynamics are more complex, and vaccination can affect host susceptibility, the infectiousness of infected hosts and the likelihood of severe disease outcomes \citep{moore2021vaccination, bouros2024prioritising}. Modelling reveals that, from the point of view of reducing transmission, optimal immunisation strategies are, in general, dependent on the underlying network topology (see, e.g., \cite{PastorSantorras2002,Cohen2003,Kitsak2010}) and, in practice, vaccine deployment is often targeted, rather than random \citep{baguelin2013assessing, bubar2021model, singer2022evaluating, jentsch2021prioritising, saadi2021models, merler2016containing}. Moreover, the primary aim of vaccination may be to reduce the prevalence of severe outcomes, rather than to reduce transmission. For example, current policy on booster vaccination against COVID-19 in England involves annual vaccination of vulnerable individuals and those aged 75 years and over, representing both a small percentage of the host population and those who typically have relatively few contacts with other individuals \citep{prem2017projecting, lovell2022estimating}.

Our modelling framework is sufficiently flexible to allow inclusion of more realistic disease dynamics and varied vaccination strategies in future. Going forwards, possible targets for further study include, among other things, exploring how our results are affected by the combined effects of multiple types of population stratification (for example, considering not only spatial structure, but also age structure \citep{nishiura2011assortativity}) or heterogeneity in infectiousness between infected individuals \citep{lloyd2005superspreading}, which could be modelled in our case by variations in the basic epidemiological parameters $\beta$ and $\gamma$. While we focussed on population vulnerability being affected by previous infections in the context of a single pathogen, in principle we could also explore the effects of cross-protection between different pathogens or multiple strains of the same pathogen \citep{thompson2019increased, thompson2023impact}; rather than a single ``recovered'' state, we could include states leading to decreased but nonzero probabilities of infection and/or transmission, with these probabilities possibly being dependent on the time after infection or vaccination. It would also be interesting to explore the effects of network architecture on the effectiveness of mitigation strategies other than vaccination, for example masking, quarantine and isolation, see for example \cite{pourbohloul2005}.

Despite the simplifications that we made and possible avenues for future work, our analyses were sufficient to demonstrate key principles about how network architecture, coupling, and the distribution of immunity in spatially structured populations affect vulnerability to outbreaks. We hope that these analyses will form the basis of future research to understand in greater detail how pathogens can spread in different geographies.

\section*{Competing Interests}
The authors declare no competing interests.

\section*{Authors' Contributions}
MU: conceptualization, methodology, formal analysis, visualisation, validation, writing – review and editing.\\
RNT: methodology, writing – review and editing.\\
MB: conceptualization, methodology, formal analysis, project administration, supervision, visualisation, writing – original draft, writing – review and editing.

\section*{AI assistance}
No AI assistance was used when carrying out or writing up the research in this manuscript, or in production of the accompanying computational resources.

\section*{Funding}
RNT acknowledges the support of the JUNIPER Consortium (MR/X018598/1). MB acknowledges support by Research England under the Expanding Excellence in England (E3) funding stream awarded to MARS: Mathematics for AI in Real-world Systems in the School of Mathematical Sciences at Lancaster University. For the purpose of open access, the authors have applied a CC BY public copyright licence to any Author Accepted Manuscript (AAM) version arising from this submission. 

\section*{Acknowledgements}
The authors thank members of the Infectious Disease Modelling group (part of the Wolfson Centre for Mathematical Biology) in the Mathematical Institute at the University of Oxford for useful discussions about this work. MB would like to thank Lloyd Chapman for helpful comments on a draft of this manuscript. All authors would like to thank the anonymous referees for a number of useful comments on this work.

\section*{Data Availability}
The computing code used to perform the analyses in this article is available, along with relevant data, in the GitHub repository at \cite{muradEpigithub}. All code was written in {\tt C/C++}.

\pagebreak

\appendix

\setcounter{figure}{0}
\makeatletter 
\renewcommand{\thefigure}{S\arabic{figure}}
\renewcommand{\theHfigure}{S\arabic{figure}}
\renewcommand{\thesection}{S\arabic{section}}
\renewcommand{\appendixname}{Appendix}
\titleformat{\section}{\normalfont\Large\bfseries}{Appendix \thesection}{1em}{}

\section*{Supporting Information}

\section{Branching process approximations and herd immunity in spatial models}
\label{appbranching}

We follow Chapter 5 in \cite{AthreyaNey}, see also \cite{AllenMBI}. A model of the form (\ref{mainreacscheme}) with $n$ subpopulations is associated with an $n$-type branching process. Formally, we obtain the corresponding branching process by assuming that the susceptible fraction in each subpopulation is fixed over the lifetime of an infected individual: a good approximation provided the numbers of susceptible individuals in each subpopulation are large. Equivalently, as an approximation to \eqref{mainreacscheme}, we consider the reaction scheme
\begin{equation}
  \label{branchingreacscheme}
  \mathsf{I}_i \overset{\varepsilon_{ij}\beta S_j/N_c}{\longrightarrow} \mathsf{I}_i + \mathsf{I}_j,\quad \mathsf{I}_i \overset{\gamma}{\longrightarrow} \emptyset\, \quad i,j = 1, \ldots, n
\end{equation}
with $S_j$ now treated as constant parameters. The formula for the contagion probability, $p_{ij}$, in \eqref{eqcontagion} follows from considering the jump chain (e.g., \cite{Norris97}) corresponding to the CTMC for \eqref{branchingreacscheme}. 

\subsection*{The mean matrix}

Define
\begin{equation}
L_i := \gamma + \beta \sum_{j=1}^n \varepsilon_{ij}\frac{S_j}{N_c} = \gamma\left(1 + \sum_{j=1}^n\wW_{ij}\right)
\end{equation}
so that $L_iI_i$ is the total propensity of reactions of \eqref{mainreacscheme} with sources including infected individuals in the $i$th subpopulation. We can then calculate the offspring probability generating functions (PGFs) for the corresponding $n$-type branching process to be
\begin{equation}
f_i(s_1, \ldots, s_n) = \frac{1}{L_i}\left[\gamma + \beta\sum_{j=1}^n \varepsilon_{ij}\frac{S_j}{N_c} s_is_j\right], \quad (i=1, \ldots, n)\,.
\end{equation}
Here $\gamma/L_i$ is the probability that a parent of type $i$ (i.e., an infected individual in subpopulation $i$) produces no offspring (i.e., recovers without infecting any other susceptible individual); while $\frac{\varepsilon_{ij}\beta S_j}{L_iN_c}$ is the probability that a parent of type $i$ produces one offsping of type $i$ and one of type $j$ (i.e., causes an infection in subpopulation $j$, where $j=i$ is possible). 

By definition, the {\em mean matrix} $M$ then has entries
\begin{equation}
  \label{eqmeanmat}
M_{ii} := \frac{\partial f_i}{\partial s_i} (\bm{1}) = \frac{\beta}{L_i}\left[ \frac{\varepsilon_{ii} S_i}{N_c} + \sum_{j=1}^n \varepsilon_{ij}\frac{S_j}{N_c}\right], \quad M_{ij} := \frac{\partial f_i}{\partial s_j} (\bm{1}) = \frac{\beta \varepsilon_{ij} S_j}{L_iN_c}\,\,\,(j \neq i)\,.
\end{equation}
Here $\bm{1}$ refers to a vector of ones. Assuming a fully susceptible population initially, we have $L_i(0) = \beta + \gamma$, and
\begin{equation}
M_{ii}(0) = \frac{\beta(\varepsilon_{ii}+1)}{\beta+\gamma}, \quad M_{ij}(0) = \frac{\beta\varepsilon_{ij}}{\beta+\gamma}\,\,\,(j \neq i)\,, \quad \mbox{so that} \quad M(0) = \frac{\beta}{\beta+\gamma}(\mathrm{Id} + \be)
\end{equation}
where $\mathrm{Id}$ is the $n \times n$ identity matrix.

\subsection*{Extinction probabilities and the herd immunity threshold}

Define $q_i$ to be the probability of eventual extinction of the $n$-type branching process associated with \eqref{mainreacscheme} given a single initial infection in subpopulation $i$; and let $q = (q_1, \ldots, q_n)$. Let us assume that the mean matrix $M$ is primitive, i.e., $M$ is nonnegative, and some power of $M$ is strictly positive. The former is automatic, and the latter is a weak assumption in practice: it is sufficient for $\be$ to be irreducible and aperiodic. Then $\rho(M)$ is, by the Perron-Frobenius theorem, a simple positive eigenvalue of $M$. Further (Thm. 2 in Chapter V.3, \citep{AthreyaNey}: (i) if $\rho(M) \leq 1$, then $q = \bm{1}$, namely, all entries of $q$ are $1$; (ii) $\lim_{k \to \infty} f^k(s) = q$ for any $s \in [0,1)^n$; and (iii) $q$ is the unique fixed point of $f:= (f_1, \ldots, f_n)$ in $[0,1)^n$. Thus we have extinction with certainty if and only if $\rho(M) \leq 1$ and, additionally, the vector $q$ of extinction probabilities is easily approximated with an iteration.

Recalling that $\wW^\top$ is the next-generation matrix, the natural definition of the {\em herd immunity threshold} ({\em HIT} for short) in a spatial model is the set of states $S:=(S_1, \ldots, S_n)$ at which the next-generation effective reproduction number is equal to $1$, namely, $\rho(\wW) = 1$, see \cite{DriesscheR0}. For the purposes of this definition we take $S \in \mathbb{R}^n_{\geq 0}$ (rather than $\mathbb{Z}^n_{\geq 0}$). Alternatively, as we show next, we may define the HIT to be the set of states at which $\rho(M)=1$.

Observe first that both $\wW$ and $M$ are nonnegative matrices, and that
\begin{equation}
\label{eqWM}
M_{ii} = \frac{\wW_{ii} + \sum_{j=1}^n\wW_{ij}}{1 + \sum_{j=1}^n\wW_{ij}} \quad \mbox{and} \quad M_{ij} = \frac{\wW_{ij}}{1 + \sum_{j=1}^n\wW_{ij}}\,.
\end{equation}
Suppose that $\rho(\wW) = 1$. Then, given any positive $z$ satisfying $\wW z=z$, $Z^{-1} \wW Z$ is a row-stochastic matrix (Thm. 5.4, Chapter 2 in \cite{berman}, for example), where $Z = \mathrm{diag}\{z_1, \ldots, z_n\}$. I.e., $(Z^{-1}\wW Z\bm{1})_i = \sum_{j=1}^n\wW_{ij}z_j/z_i = 1$, for each $i = 1, \ldots, n$. It follows, using \eqref{eqWM}, that
\begin{equation}
(Z^{-1}MZ\bm{1})_i = \frac{\wW_{ii} + \sum_{j=1}^n\wW_{ij}}{1 + \sum_{j=1}^n\wW_{ij}} + \frac{\sum_{j=1,j\neq i}^n\wW_{ij}z_j/z_i}{1 + \sum_{j=1}^n\wW_{ij}} = \frac{\sum_{j=1}^n\wW_{ij}z_j/z_i + \sum_{j=1}^n\wW_{ij}}{1 + \sum_{j=1}^n\wW_{ij}} = 1\,.
\end{equation}
Thus $Z^{-1}MZ$ is row-stochastic, and $\rho(M)=1$. Similarly, if $\rho(M) = 1$, then $\rho(\wW) = 1$.

We close this section by noting that, in the stochastic setting, crossing the HIT corresponds to the (major) outbreak probability becoming zero, provided populations are large and so the hypotheses of the branching-process approximation hold. However, for smaller subpopulations (including the extreme case of individual-based networks), the branching process approximation breaks down, and the theoretical herd immunity threshold overestimates the levels of prior immunity required to make outbreak probabilities effectively zero.

\newpage 
\section{Analytical estimates for complete, symmetric networks}
\label{appcomplete}
For a complete, symmetric network, the matrix $\wW$ corresponding to (\ref{mainreacscheme}) has entries
\begin{equation}
  \label{nextgensym}
\wW_{ii} = \frac{(1-(n-1)\varepsilon)\beta S_i}{\gamma N_c} = (1-(n-1)\varepsilon)s_i\wR_0, \quad \wW_{ij} = \frac{\varepsilon \beta S_j}{\gamma N_c} = \varepsilon s_j \wR_0\,\,\,(i \neq j)\,,
\end{equation}
where $\varepsilon := \varepsilon_{ij}$ is the same for all $i \neq j$, and $s_i$ is the susceptible fraction in subpopulation $i$. The mean leak is now
\begin{equation}
\Theta = (n-1) \varepsilon\,.
\end{equation}

\subsection*{Contagion probabilities}
Substituting the expressions from (\ref{nextgensym}) into (\ref{eqalpha1}), for sufficiently small $\varepsilon \wR_0$, the probability an outbreak of size $m$ in subpopulation $i$ {\em fails} to trigger an outbreak in subpopulation $j$ is approximately
\begin{equation}
\left(1-\varepsilon\left(s_j\wR_0-\frac{1}{1-(n-1)\varepsilon}\right)\right)^m.
\end{equation}
Recalling (\ref{lambertmain}), which gives $E(s, \wR_0)$, the expected fraction infected, given an outbreak in a subpopulation with susceptible fraction $s$, we can then define
\begin{equation}
  \label{alphageneral}
\alpha(\varepsilon, s_i, s_j, \wR_0, N_c):=\left(1-\varepsilon\left(s_j\wR_0-\frac{1}{1-(n-1)\varepsilon}\right)\right)^{N_cE(s_i,\wR_0)}\,,
\end{equation}
and observe that $1-\alpha(\varepsilon, s_i, s_j, \wR_0, N_c)$ is the approximate probability of contagion from subpopulation $i$ to subpopulation $j$, given an outbreak in subpopulation $i$. 

\subsection*{Expected outbreak sizes}
Let $p_{n,m}(\alpha)$ be the probability that an outbreak in a complete, symmetric, $n$-subpopulation model with contagion probability $1-\alpha$ hits exactly $m \leq n$ subpopulations. Clearly $p_{1,1}=1$ and, by an easy combinatorial argument, we can show that $p_{n+1,m}(\alpha) = n\alpha^mp_{n,m}(\alpha)/(n-m+1)$, allowing us to compute $p_{n,m}(\alpha)$ for all $n$ and all $m \leq n$.

Now suppose all subpopulations have the same susceptible fraction, let $s:=s_1 = s_2 = \cdots = s_n$, and write $\alpha =\alpha(\varepsilon, s, s, \wR_0, N_c)$ for brevity. Recalling (\ref{lambertmain}), the expected outbreak size, as a fraction of the total population, conditioned on the occurrence of an outbreak, is then, approximately,
\begin{equation}
  \label{alphapoly}
\frac{E(s, \wR_0)}{n}\sum_{m=1}^nmp_{n,m}(\alpha)\,.
\end{equation}
Multiplying through by the approximate probability of an outbreak occurring in any given subpopulation given an introduction, namely $1-1/(s\wR_0)$, gives us, approximately, the (unconditional) expected fraction infected given a random introduction, namely
\begin{equation}
  \label{alphapoly1}
\frac{E(s, \wR_0)(s\wR_0-1)}{ns\wR_0}\sum_{m=1}^nmp_{n,m}(\alpha)\,.
\end{equation}
Computing $\alpha$ from (\ref{alphageneral}) and substituting into (\ref{alphapoly1}) we obtain estimates of the outbreak size as we vary either the mean leak (and hence $\varepsilon$), or the level of prior random immunity (hence $s$) in complete, symmetric models with weak coupling and large subpopulations. These calculations are used to derive the theoretical curves in Figure~\ref{fig_leak1} (left)~and~Figure~\ref{fig_vax10c} (left).

\newpage
\section{Outbreak size simulations}
\label{appsim}

All numerical results presented in this paper are based on exact stochastic simulations using the Gillespie algorithm \citep{Gillespie}, carried out in C/C++ \citep{muradEpigithub} on a personal computer. The commands used to run all the simulations, and produce all the figures, in this manuscript are available on the associated GitHub site in the file {\tt SimulationNotes}. All plots and videos are also available on the GitHub site.

{\bf Ensuring connected networks.} All networks simulated are connected: if a random procedure, such as constructing a random network, or rewiring a lattice, results in a disconnected network, we discard the network and try again. 

{\bf Random introductions and immunisations.} In simulations, assuming a nonzero susceptible population, when we make a random introduction, we randomly choose a subpopulation, say subpopulation $i$, and in order to maintain constant total population, we increment $I_i$ and decrement $S_i$; if $S_i=0$, we randomly choose another subpopulation and try again until we have successfully increased the total number of infections. Immunising an individual in subpopulation $i$ means decrementing $S_i$ and incrementing $R_i$. The probability that a random immunisation occurs in subpopulation $i$ is proportional to the fraction of the total susceptible population in subpopulation $i$. 

{\bf Mean outbreak sizes.} With the exception of the simulations leading to Figure~\ref{figRt}, mean outbreak sizes are always computed from a single random introduction into a population with no other current infections, using a varying number $n_{sim}$ of stochastic simulations, dependent on the network size and the complexity, and hence the associated computational cost of each simulation. To compute mean outbreak sizes along a disease trajectory in Figure~\ref{figRt} we first project the current state onto the disease-free states by setting all infected individuals to be recovered, and then carry out the above procedure. 

{\bf The effects of random immunisation on outbreak size.} The following procedure is used in the simulations leading to Figure~\ref{fig_vax10c} and Figure~\ref{fig_vaxreversed}. We choose integers $n_1$ and $n_2$. For each level of immunisation we carry out $n_1$ sets of random immunisations, and then carry out $n_{sim} = n_2$ simulations to determine the mean outbreak size for each such immunisation. The plotted value is the mean of these means. For random and ring-preferential networks we set $n_1=10$ and $n_2 = 1000$. In all other cases we set $n_1 = 1$ and $n_2 = 10000$. 

{\bf The effects of architecture on outbreak size.} The following procedure is used in the simulations leading to Figures~\ref{figarchitecture},~\ref{figarchitecture1}~and~\ref{figarchitecture2}. We choose integers $n_1$ and $n_2$. For each parameter value which controls network architecture, we carry out $n_{sim} = n_2$ simulations to determine the mean outbreak size in each of $n_1$ randomly chosen networks. The resulting mean outbreak size for each network is shown as a blue dot; the mean of all means is shown as a red dot. In all simulations shown, $n_1 = 100$ and $n_2 = 1000$. While $\wR_0$ is held constant at $2$ in all of these simulations, other parameters, including the mean leak and certain parameters determining network structure, are chosen to ensure that the maximal outbreak sizes across the different classes of networks are similar.

{\bf The effects of architecture on outbreak size distributions.} Each histogram of outbreak sizes shown in Figure~\ref{figsizehist} is generated from 50000 outbreak simulations, and a bin size of 2\% is used. Outbreaks affecting less than 2\% of the population are omitted from the plots so the distribution can be visualised more easily.

{\bf Comparisons between the protective effects of infection-acquired and random immunity.} The following procedure is used in the simulations leading to Figure~\ref{fig_protectmain} and Figures~\ref{fig_protectrandindividual}--\ref{fig_protectmainalt1}. We choose integers $n_1$ and $n_2$. We initiate primary outbreaks through a random introduction, and let them run their course. Let us suppose that a given outbreak ends with $d\%$ immunity. To ensure that we obtain data across a range of primary outbreak sizes we choose the first 10 simulations where $d$ falls into each interval: $1-5$, $5-10$, $10-15$, etc. Note that the first interval is chosen to start at $1\%$ to avoid picking up a large number of minor outbreaks. Also note that for some networks with strongly bimodal outbreak size distributions, achieving 10 primary outbreaks in each bin may require numerous simulations. In order to achieve a range of primary outbreak sizes, we are forced to avoid parameters which lead to very strongly bimodal outbreak-size distributions.

Following each primary outbreak simulation ending with $d\%$ immunity in the population:
\begin{enumerate}
\item Starting from the final state after the primary outbreak, we find the mean size of secondary outbreaks (blue circles) using $n_{sim} = n_1n_2$ new simulations. 
\item Starting from the fully susceptible state, we introduce $d\%$ random immunity $n_1$ times, and for each of these initial conditions we estimate mean outbreak size using $n_{sim} = n_2$ simulations. The mean of these $n_1$ means gives a (brown) square, which we connect via a line to the blue circle for comparison.
\end{enumerate}
For most simulations, $n_1 = 10$ and $n_2=1000$. However, for some simulations, especially those with many small subpopulations, more simulations are needed; the details are in the file {\tt SimulationNotes} on the accompanying GitHub \citep{muradEpigithub}. 

Note that in these simulations, we only check secondary outbreak size at the end of the primary outbreak; i.e., we never stop unfinished outbreaks and project onto the disease-free states. 

{\bf The fraction of outbreaks leading to herd immunity.} In generating the data in Table~\ref{table_herd}, we first compute mean outbreak size using $n_{sim} = 10000$ simulations. We then fix a threshold $h$ for ``major outbreaks'' (see Section~\ref{secprotect}) and generate $1000$ primary outbreaks. Whenever a primary outbreak is major, we compute, as above, the mean secondary outbreak size using $n_{sim}$ simulations. If the computed mean secondary outbreak size passes the threshold $h$, we say that herd immunity was not reached after the primary outbreak. Note that if we decrease $h$, we expect the fraction of major outbreaks leading to herd immunity to decrease.

\newpage
\section{Additional simulation results}
\label{secgraphs}

\subsection*{The sigmoidal response to increasing leak}

\begin{figure}[htp!]
\begin{center}
  \underline{The response of normalised outbreak size to mean leak across architectures}

  \includegraphics[scale=0.4]{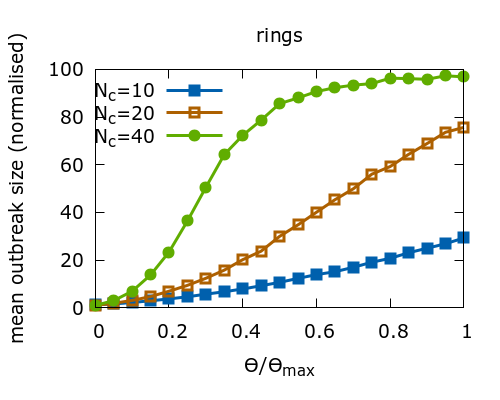}
  \includegraphics[scale=0.4]{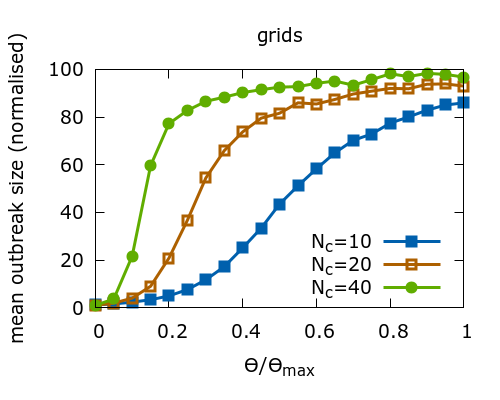}
  \includegraphics[scale=0.4]{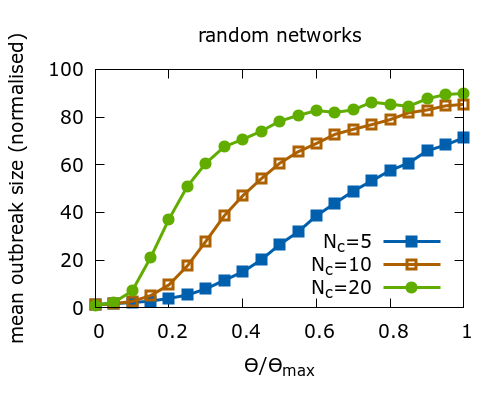}
  \includegraphics[scale=0.4]{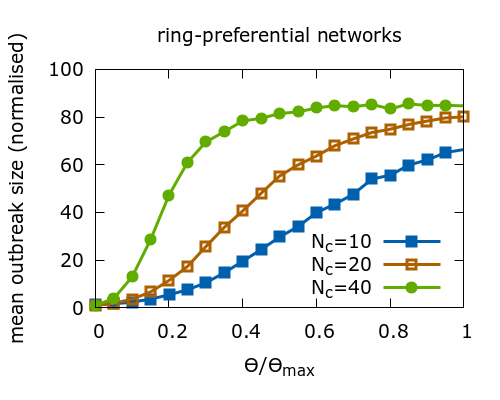}
\end{center}

\caption{\label{figsigmoidal}As we increase $N_c$ we see the characteristic sigmoidal response of (normalised) outbreak sizes to leak. {\em Top left.} Rings with $n=100$, $K=4$, no rewiring, and $N_c = 10, 20$ and $40$. {\em Top right.} $10 \times 10$ grids with no rewiring, and $N_c = 10, 20$ and $40$. {\em Bottom left.} Random networks with $n=100$, expected mean degree $\hat{d}=4$ and $N_c = 5, 10$ and $20$. {\em Bottom right.} Ring-preferential networks with $n=100$, initial ring size $10$, initial clique size $2$, attachment degree $1$ and $N_c = 10, 20$ and $40$.}
\end{figure}

In Figure~\ref{fig_leak1}, we saw examples of the monotonic response of mean outbreak size to increasing leak. In Figure~\ref{figsigmoidal} we demonstrate that, across a range of network architectures, as we increase subpopulation size, the response of mean outbreak size to mean leak becomes more clearly sigmoidal. In other words, in models with large subpopulations, regardless of architecture, there is some threshold level of transmission between subpopulations near which the mean outbreak size changes rapidly. 

\newpage
\subsection*{Network architectures and outbreak sizes}

\begin{figure}[htp!]
\begin{center}
  \underline{Outbreak size as we vary network architecture in constant-infectivity networks}

  \includegraphics[scale=0.4]{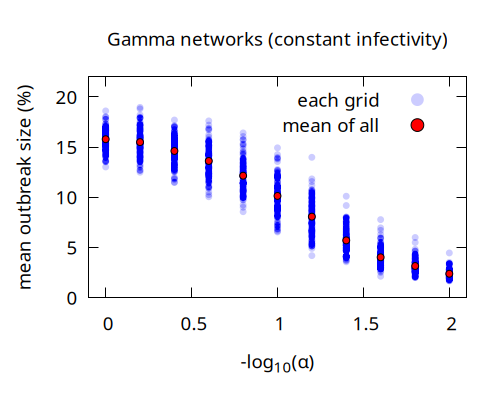}
  \includegraphics[scale=0.4]{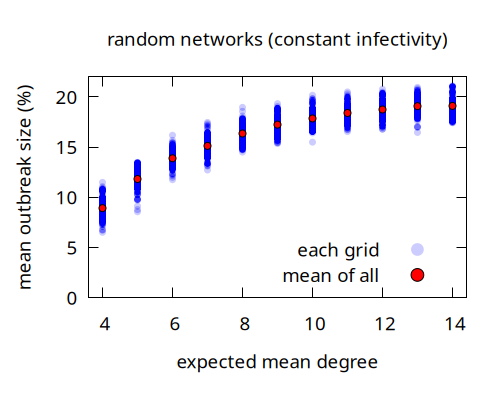}
  \includegraphics[scale=0.4]{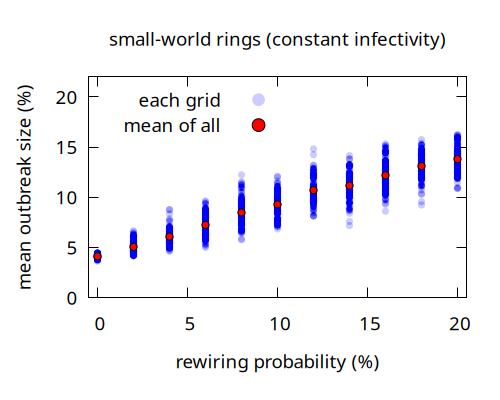}
  \includegraphics[scale=0.4]{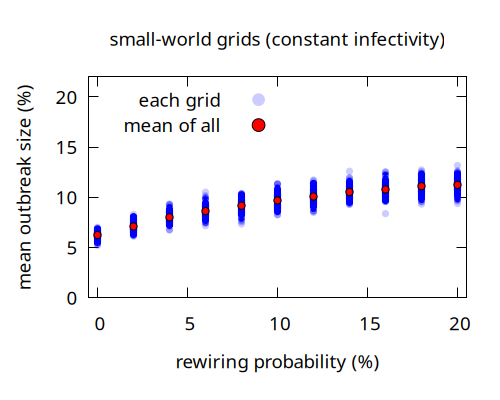}
  
\end{center}

\caption{\label{figarchitecture1}For each parameter value, random introductions to $100$ randomly chosen networks were simulated $n_{sim}=1000$ times each to determine the mean outbreak size (blue dots), with the global mean shown as a red dot. All parameters are as in Figure~\ref{figarchitecture}, except that in the random networks we set $\Theta N_c = 3$.
}
\end{figure}

\begin{figure}[htp!]
\begin{center}
  \underline{Outbreak size as we vary network architecture in constant-susceptibility networks}
        
  \includegraphics[scale=0.4]{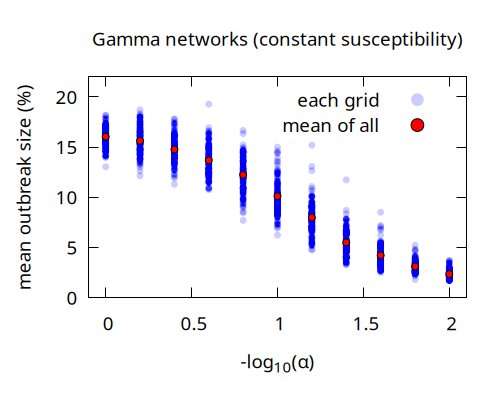}
  \includegraphics[scale=0.4]{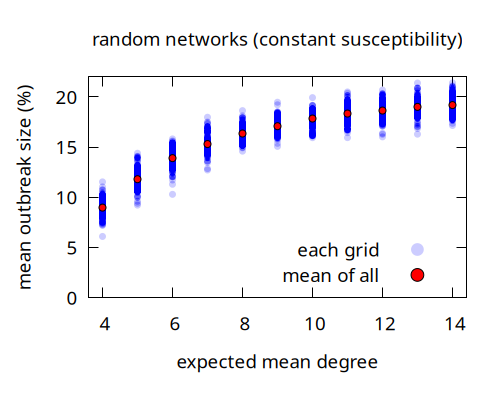}
  \includegraphics[scale=0.4]{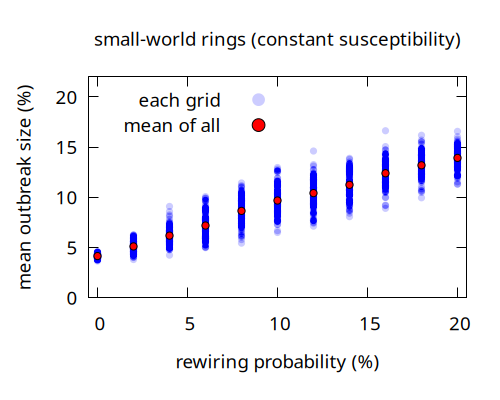}
  \includegraphics[scale=0.4]{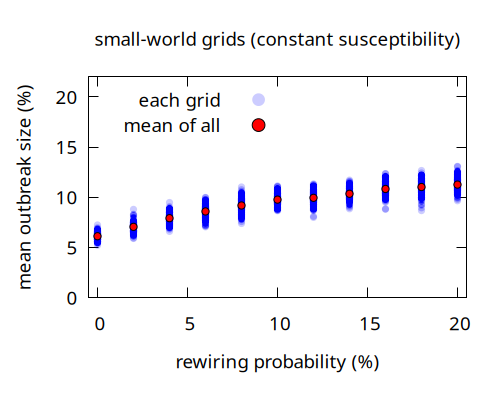}
  
\end{center}

\caption{\label{figarchitecture2}For each parameter value, random introductions to $100$ randomly chosen networks were simulated $n_{sim}=1000$ times each to determine the mean outbreak size (blue dots), with the global mean shown as a red dot. All parameters are as in Figure~\ref{figarchitecture1}.
}
\end{figure}

\begin{figure}[htp!]
  \begin{center}
\underline{Outbreak size distribution in small-world grids with increasing rewiring}
    
  \includegraphics[scale=0.4]{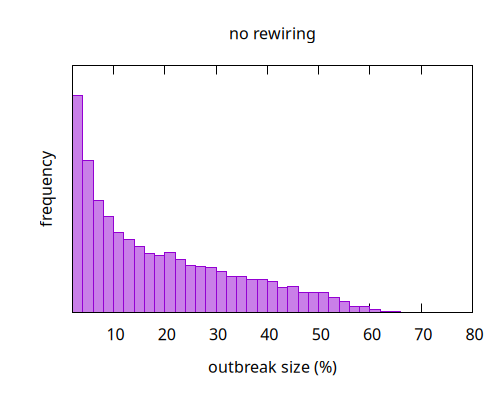}
  \includegraphics[scale=0.4]{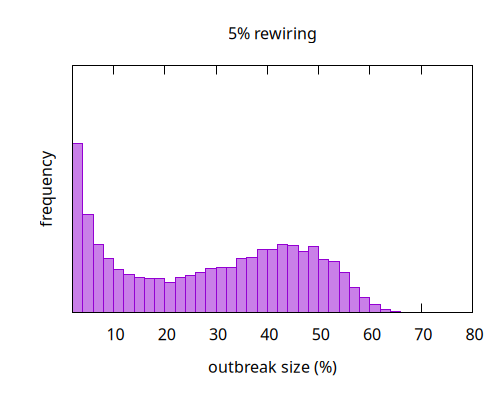}
  \includegraphics[scale=0.4]{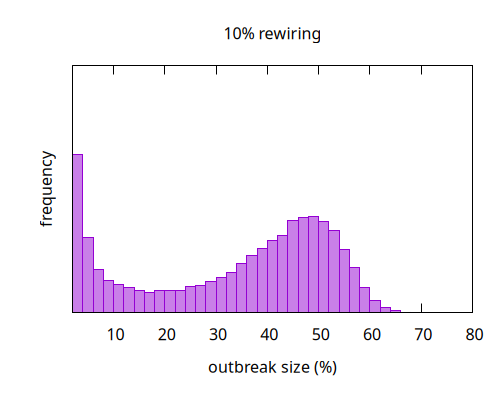}
  \includegraphics[scale=0.4]{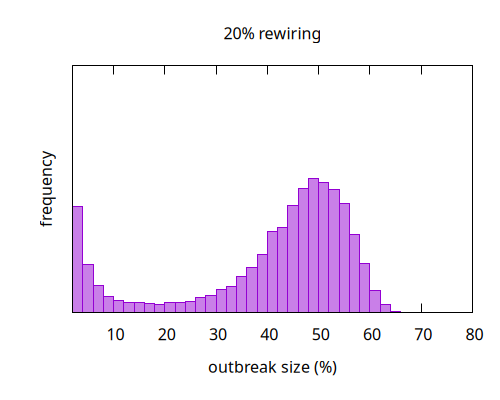}
  
\end{center}

\caption{\label{figsizehist}Distribution of outbreak sizes in small-world $19 \times 13$ grids with $N_c = 30$, $\Theta N_c = 3.5$ (as in Figure~\ref{figarchitecture}(d)), and variable rewiring probability. Each histogram is based on 50000 simulations, with outbreaks of size less than $2\%$ omitted. We see that as the rewiring probability increases, the distribution of outbreak sizes becomes more clearly bimodal.
  }

\end{figure}

In Figure~\ref{figarchitecture}, we saw the effects of network architecture on mean outbreak sizes in default networks. The response of the mean outbreak size as we vary network architecture in constant-infectivity and constant-susceptibility networks is shown in Figures~\ref{figarchitecture1}~and~\ref{figarchitecture2}. The response of outbreak size to architecture is qualitatively very similar in constant-infectivity and constant-susceptibility networks to the response in the default case. In Figure~\ref{figsizehist} we see how, as the rewiring probability increases in small-world grids, the distribution of outbreak sizes becomes more clearly bimodal.

\newpage

\newpage

\newpage
\subsection*{Prior random immunity in spatial models can sometimes be less protective than in homogeneous models}

\begin{figure}[h]
\begin{center}
  \underline{Outbreak sizes versus prior random immunity in random and ring-preferential networks}
  
  \includegraphics[scale=0.37]{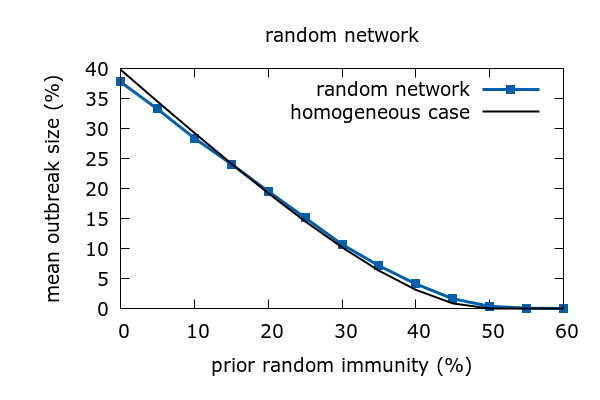}  
  \includegraphics[scale=0.37]{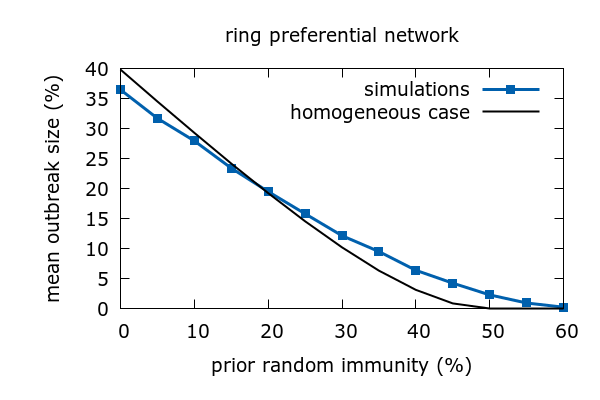}
\end{center}

\caption{\label{fig_vaxreversed}Mean outbreak sizes as a function of prior random immunity in random and ring-preferential networks with strong coupling. In both cases, $n=20$, $N_c=1000$ and $\Theta N_c = 667$. {\em Left.} A random network with expected mean degree $\hat{d} = 6$ (so $\Theta \approx 0.78\Theta_{max}$). {\em Right.} A ring-preferential network with initial ring size 4, clique size $2$, and attachment degree $1$ (so $\Theta \approx \Theta_{max}$).}
\end{figure}

The simulations in Figure~\ref{fig_vax10c} demonstrated that prior random immunity often offers greater protection to a spatially structured population than to the corresponding homogeneous population. However, the benefits of spatial structure can disappear, or even be reversed, for some architectures when the coupling between subpopulations is sufficiently strong. In Figure~\ref{fig_vaxreversed}, we see that in a strongly coupled random network, spatial structure offers no advantage; while in a strongly-coupled ring-preferential network, random immunity can in fact provide less protection than in the homogeneous case.

\newpage
\subsection*{Infection-acquired versus random immunity: effects of subpopulation size}

We examine how varying subpopulation size $N_c$ affects whether infection-acquired or random immunity is more protective in random and ring-preferential networks. As $N_c$ increases, corresponding to greater spatial clustering of the population, we see that random immunity becomes more protective than infection-acquired immunity.

\begin{figure}[htp!]
\begin{center}
  \underline{Outbreak size versus prior immunity in random networks with varying subpopulation size}
    
  \includegraphics[scale=0.35]{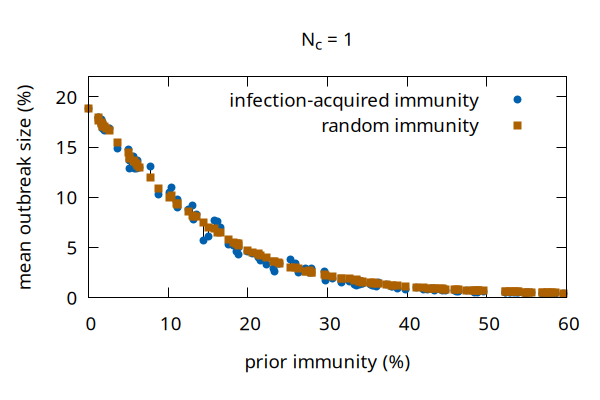}
  \includegraphics[scale=0.35]{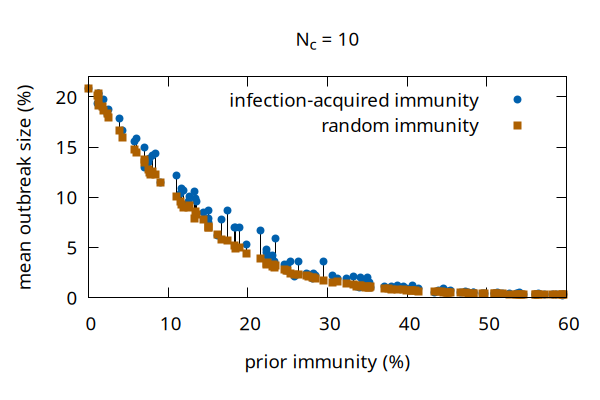}
  \includegraphics[scale=0.35]{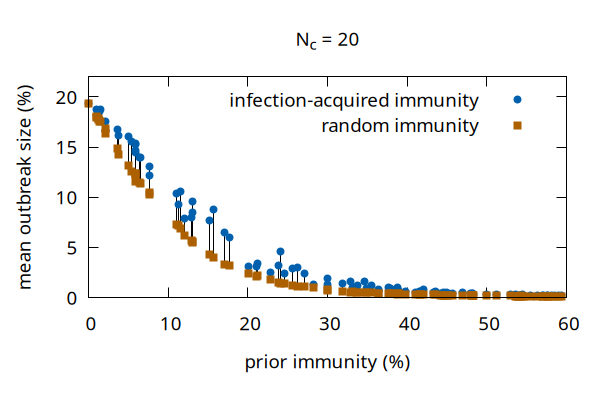}
  \includegraphics[scale=0.35]{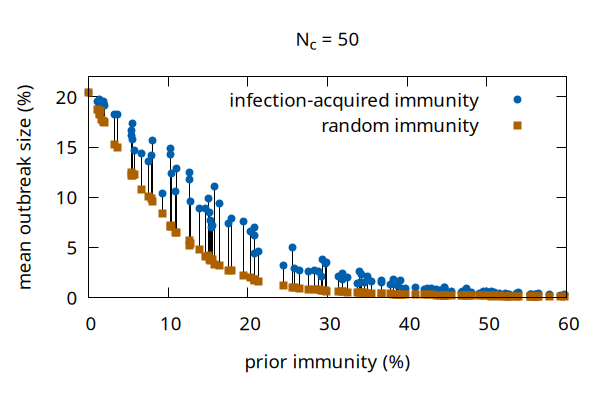}
\end{center}

\caption{\label{fig_protectrandindividual}Mean outbreak sizes following an outbreak, or following addition of the same amount of random immunity, in random networks. In all cases $\hat{d} = 5$ and, with the exception of the individual-based network, $\Theta N_c = 3.5$ (in the case of individual-based networks, recall that $\Theta$ always takes the value $1$). The number of subpopulations $n$ is chosen so that the networks have similar mean outbreak sizes in the absence of prior immunity. {\em Top left.} $n=500$, $N_c = 1$ (i.e., 500 individuals). {\em Top right.} $n=100$, $N_c=10$. {\em Bottom left.} $n=100$, $N_c = 20$. {\em Bottom right.} $n=50$, $N_c=50$. There is no discernable difference in the protective effects of random immunity and infection-acquired immunity in the individual-based network; however, random immunity becomes clearly more protective than infection-acquired immunity as the subpopulation size $N_c$ increases.
}
  
\end{figure}

The results for random networks with varying subpopulation size $N_c$ are shown in Figure~\ref{fig_protectrandindividual}. We see that in the individual-based network infection-acquired and random immunity provide roughly the same level of protection; but as subpopulation size increases random immunity becomes clearly more protective than infection-acquired immunity.

\begin{figure}[htp!]
  \begin{center}
  \underline{Outbreak size versus prior immunity in ring-preferential networks with varying subpopulation size}

    \includegraphics[scale=0.35]{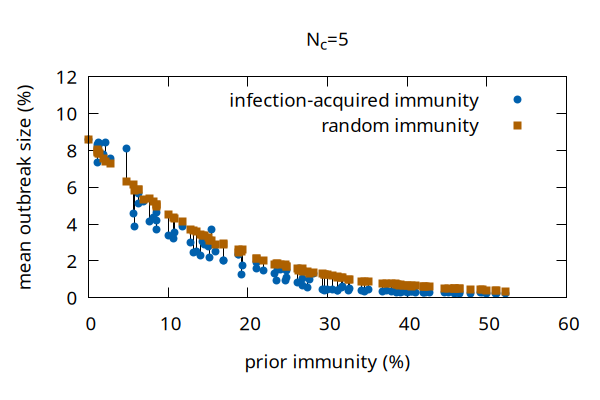}
    \includegraphics[scale=0.35]{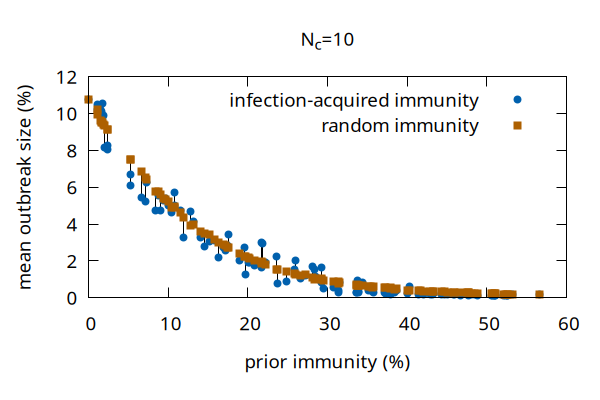}
    \includegraphics[scale=0.35]{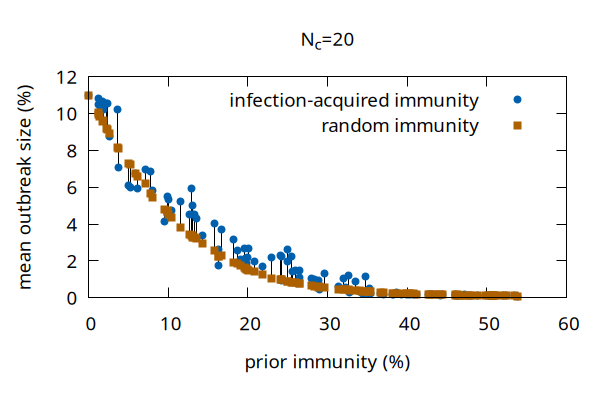}
    \includegraphics[scale=0.35]{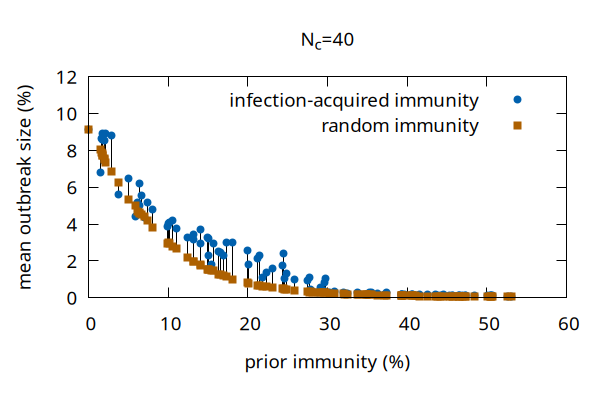}
\end{center}

  \caption{\label{figringpref}Mean outbreak sizes following an outbreak, or following addition of the same amount of random immunity, in ring-preferential networks. All networks have $n=200$, initial ring size $20$, clique size $4$, and attachment degree $1$. Mean leak is set to achieve comparable baseline outbreak sizes in each case. {\em Top left.} $N_c=5, \Theta N_c=2$, {\em Top right.} $N_c=10, \Theta N_c=2.5$, {\em Bottom left.} $N_c=20, \Theta N_c=3$, {\em Bottom right.} $N_c=40, \Theta N_c=3$. For the networks with smaller subpopulations, infection-acquired immunity is more protective than random immunity; however this is reversed as subpopulation size increases.
}
\end{figure}

\newpage
In Figure~\ref{figringpref} we see that as we increase subpopulation size in ring-preferential networks, while maintaining outbreak sizes, we reverse the trend that infection-acquired immunity is more protective than random immunity.

\newpage
\subsection*{Infection-acquired versus random immunity: individual based grids}
\begin{figure}[htp!]
  
\begin{center}
\underline{Outbreak size versus prior immunity in individual-based small-world grids}

  \includegraphics[scale=0.35]{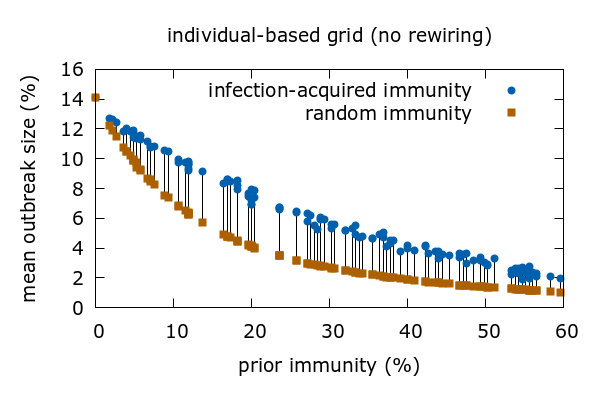}
  \includegraphics[scale=0.35]{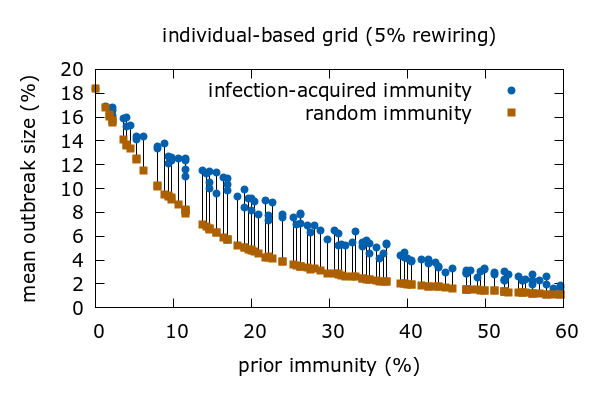}
  \includegraphics[scale=0.35]{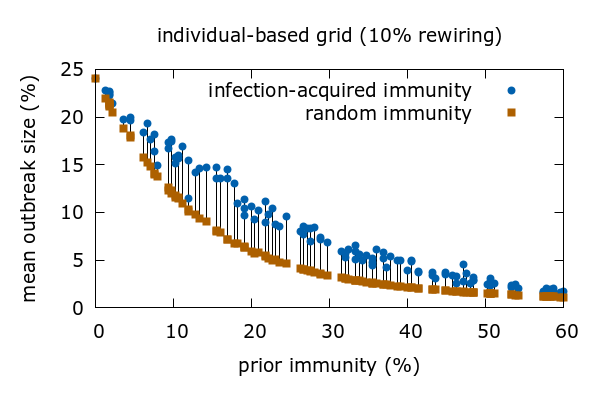}
  \includegraphics[scale=0.35]{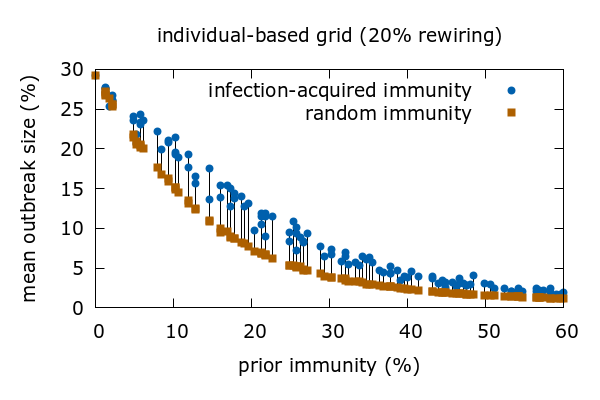}
\end{center}

\caption{\label{fig_protectgridindividual}Mean outbreak sizes following an outbreak, or following addition of the same amount of random immunity, in individual-based grids as we increase the rewiring probability $p$. All networks begin with a $15 \times 15$ grid with $\wR_0 = 3$, and we increase $p$ from $0\%$ to $5\%$, to $10\%$, to $20\%$. As expected, outbreak sizes increase with $p$, but the conclusion that random immunity is more protective than infection-acquired immunity is robust to increases in $p$ in this range.
}
  
\end{figure}

Many of the conclusions of this paper extend to individual-based networks. In Figure~\ref{fig_protectgridindividual}, we examine individual-based grids, and find that as we increase the rewiring probability, outbreak sizes increase substantially, but random immunity remains more protective than infection-acquired immunity.

\newpage
\subsection*{Infection-acquired versus random immunity: individual based ring-preferential networks}
\begin{figure}[htp!]
\begin{center}
  \underline{Outbreak size versus prior immunity in individual-based ring-preferential networks}

  \includegraphics[scale=0.35]{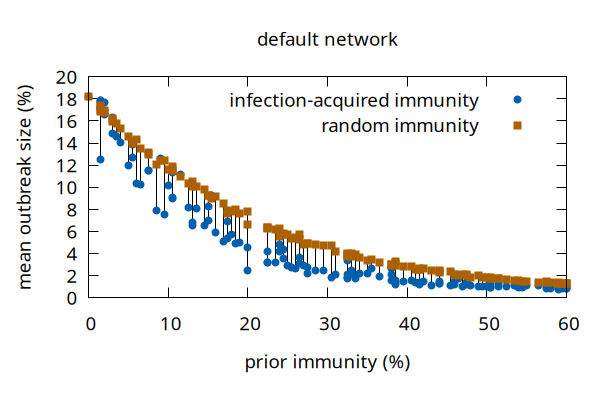}
  \includegraphics[scale=0.35]{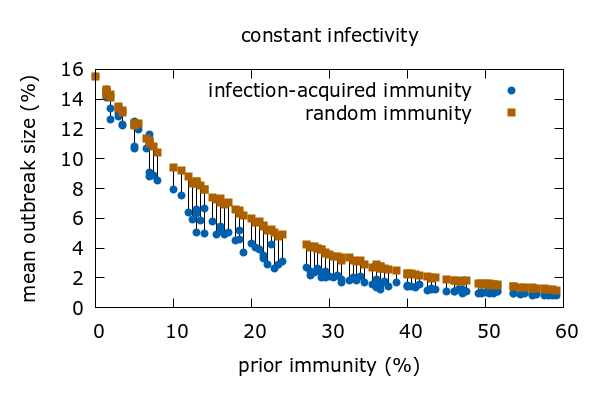}
  \includegraphics[scale=0.35]{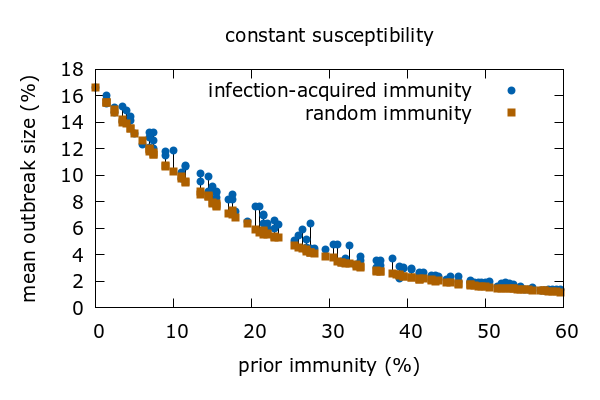}
  \includegraphics[scale=0.35]{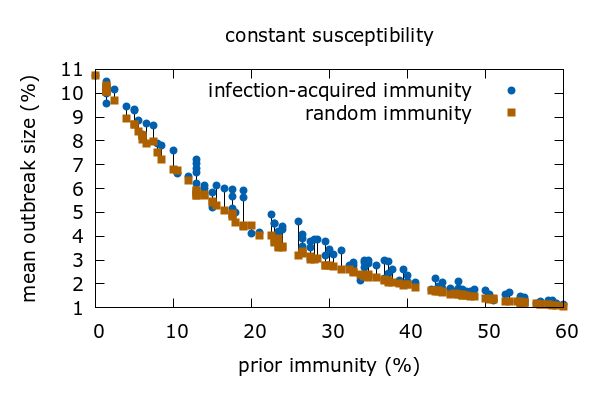}
\end{center}

\caption{\label{fig_protectrsfindividual}Mean outbreak sizes following an outbreak, or following addition of the same amount of random immunity, in individual-based, ring-preferential networks. All networks have $n=200$ and initial ring size of $10$. In the first three cases, other parameters are chosen to ensure approximately similar mean outbreak sizes in the absence of prior immunity. {\em Top left.} Default network with initial clique size and attachment degree $2$. {\em Top right.} Constant-infectivity network with initial clique size and attachment degree $4$. {\em Bottom left.} Constant-susceptibility network with initial clique size and attachment degree $4$. {\em Bottom right.} Constant-susceptibility network with initial clique size $6$ and attachment degree $3$. The simulations demonstrate that network frailty effects appear to be lost in constant-susceptibility networks (although not in constant-infectivity networks).}

\end{figure}

In Figure~\ref{fig_protectrsfindividual} we examine individual-based ring-preferential networks. We see that the conclusion that infection-acquired immunity is more protective than random immunity in these networks is reversed if we insist on constant susceptibility, but not if we insist on constant infectivity. This suggests that it is large variations in susceptibility which underlie infection-acquired immunity being more protective than random immunity. 

\newpage
\subsection*{Infection-acquired versus random immunity in constant-infectivity networks} 
\begin{figure}[htp!]
\begin{center}
  \underline{Outbreak size versus prior immunity in constant-infectivity networks}

  \includegraphics[scale=0.35]{pics/complete_2.0_2.5_50_1_30.png}
  \includegraphics[scale=0.35]{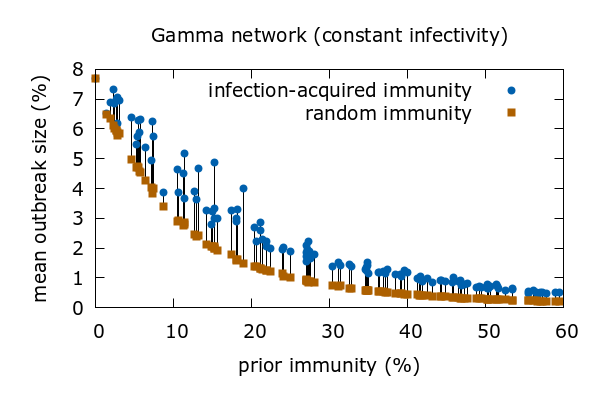}
  \includegraphics[scale=0.35]{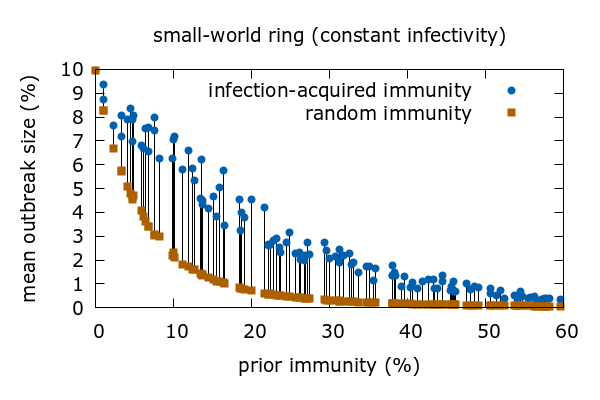}
  \includegraphics[scale=0.35]{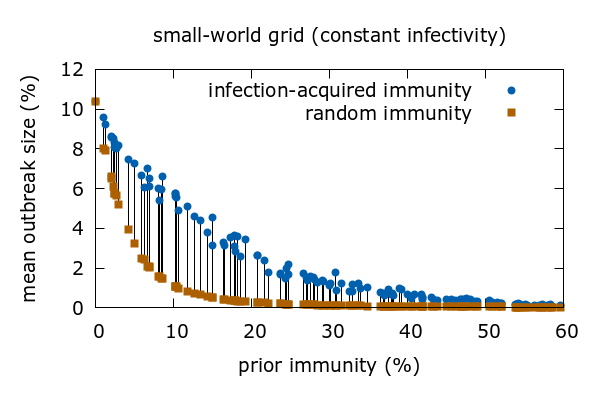}
  \includegraphics[scale=0.35]{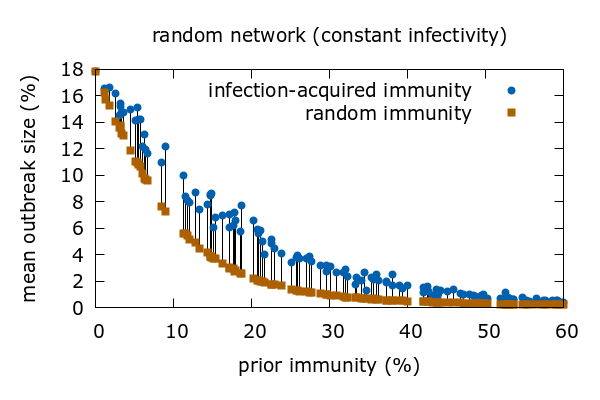}
  \includegraphics[scale=0.35]{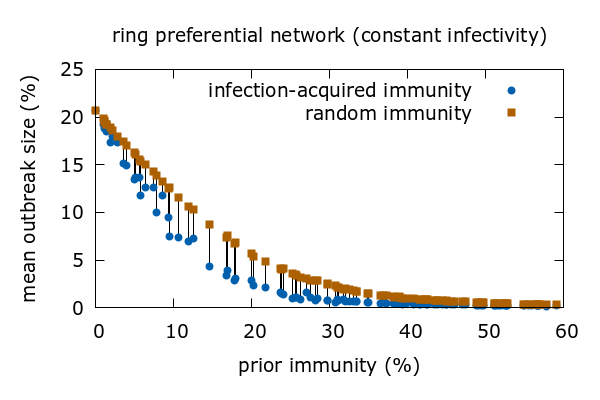}
\end{center}

  \caption{\label{fig_protectmainalt}Mean outbreak sizes following an outbreak, or following addition of the same amount of random immunity, in {\em constant-infectivity} networks. The first plot is simply reproduced from Figure~\ref{fig_protectmain}. All network parameters are as in Figure~\ref{fig_protectmain}, except in the case of the ring-preferential network (bottom right), which has $n=200$, $N_c = 5$, $\Theta N_c=4.0$, initial ring size 10, clique size $6$, and attachment degree $3$ (constant-infectivity ring preferential networks with the parameters in Figure~\ref{fig_protectmain} admit only very small outbreaks.)}
  
\end{figure}

In Figure~\ref{fig_protectmainalt}, we show that the main conclusion that for many network architectures infection-acquired immunity is less protective than random immunity, as demonstrated in Figure~\ref{fig_protectmain}, continues to hold for constant-infectivity networks.

\newpage
\subsection*{Infection-acquired versus random immunity in constant-susceptibility networks}
\begin{figure}[htp!]
\begin{center}
  \underline{Outbreak size versus prior immunity in constant-susceptibility networks}

  \includegraphics[scale=0.35]{pics/complete_2.0_2.5_50_1_30.png}
  \includegraphics[scale=0.35]{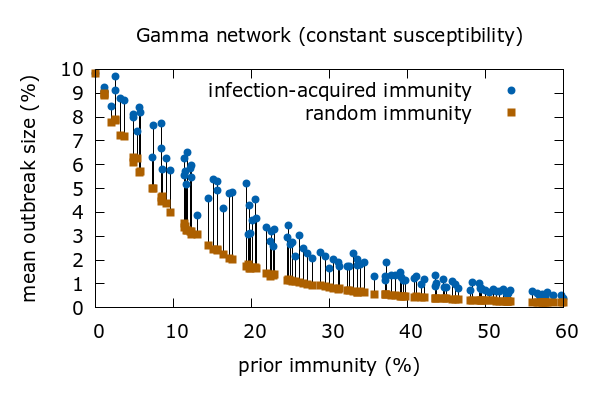}
  \includegraphics[scale=0.35]{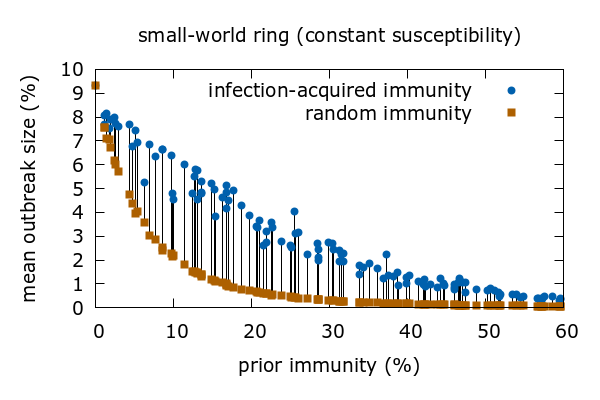}
  \includegraphics[scale=0.35]{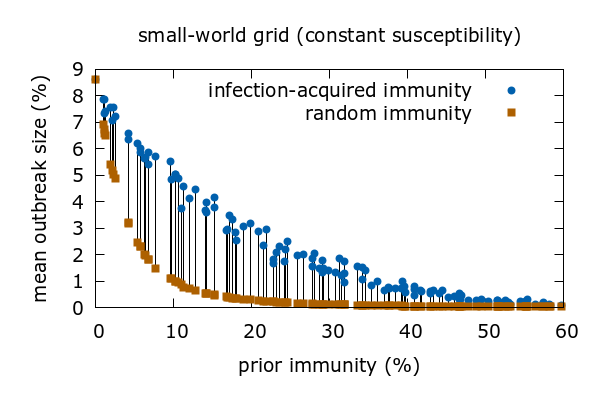}
  \includegraphics[scale=0.35]{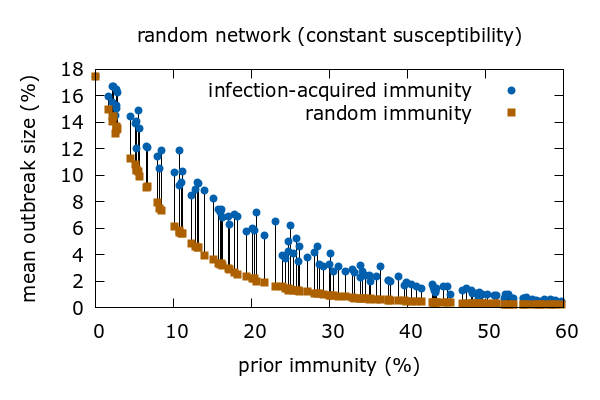}
  \includegraphics[scale=0.35]{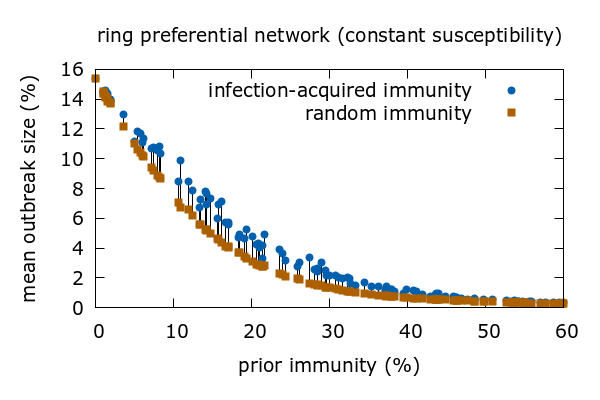}
\end{center}

  \caption{\label{fig_protectmainalt1}Mean outbreak sizes following an outbreak, or following addition of the same amount of random immunity, in {\em constant-susceptibility} networks. All network parameters are as in Figure~\ref{fig_protectmainalt}.}

\end{figure}

In Figure~\ref{fig_protectmainalt1}, we see that for constant-susceptibility networks, the conclusion that infection-acquired immunity is less protective than random immunity extends to ring-preferential networks, suggesting that large variations in susceptibility are key to network frailty effects.

\newpage

\bibliographystyle{plainnat}

\end{document}